\documentclass[journal]{IEEEtran}
\usepackage{cite}
\usepackage{tabularx,booktabs}
\usepackage{multirow}
\usepackage{subcaption}
\usepackage{amsthm}

\theoremstyle{remark}
\newtheorem{remark}{Remark} % Standard numbered remark
\newtheorem*{remark*}{Remark} % Unnumbered remark
\usepackage{amsmath,amssymb,amsfonts}
\usepackage{algorithmic}
\usepackage{graphicx}
\usepackage{textcomp}
\usepackage{xcolor}
\usepackage{mathtools}
\newcolumntype{Y}{>{\centering\arraybackslash}X}
\def\BibTeX{{\rm B\kern-.05em{\sc i\kern-.025em b}\kern-.08em
    T\kern-.1667em\lower.7ex\hbox{E}\kern-.125emX}}

\begin{document}

\title{Limiting the Impact of AI Data Centers on Fatigue Life of Thermal Turbine Generators in the Grid: A Frequency-Domain Approach }

% \author{\IEEEauthorblockN{Fiaz Hossain, Nilanjan Ray Chaudhuri, Alok Sinha, Sai Gopal Vennelaganti, and Mohammed E. Nassar}
% \IEEEauthorblockA{\textit{School of Electrical Engineering and Computer
% Science}, The Pennsylvania State University, University Park, PA, USA \\
% emails: fbh5142@psu.edu, nuc88@psu.edu, axs22@psu.edu, svennelaganti@tesla.com
% }}
\author{Fiaz Hossain,~\IEEEmembership{Student Member,~IEEE},  Nilanjan Ray Chaudhuri,~\IEEEmembership{Senior Member,~IEEE}, Alok Sinha,\\ Sai Gopal Vennelaganti,~\IEEEmembership{Member,~IEEE}, and Mohammed E. Nassar,~\IEEEmembership{Senior Member,~IEEE}
\thanks{F. Hossain and N. R. Chaudhuri are with The School of Electrical Engineering and Computer
Science, Penn State University, University Park, PA, USA e-mail: fbh5142@psu.edu, nuc88@psu.edu.}
\thanks{A. Sinha is with Department of Mechanical Engineering, Penn State University, University Park, PA, USA e-mail: axs22@psu.edu.}
\thanks{S. G. Vennelaganti and M. E. Nassar are with Tesla Inc., Palo Alto, CA, USA e-mail: svennelaganti@tesla.com, monassar@tesla.com.}
\thanks{Financial support from Tesla Inc.  under agreement 304869 PSNDA-0003632
 is gratefully acknowledged.}}

\maketitle 

\begin{abstract}
%A frequency domain based approach is proposed to impose optimal limits on artificial intelligence (AI) data center (DC) load variations so that the steam/gas turbine fatigue life of the synchronous generators (SGs) in the system is not negatively impacted. %To make the approach easily accessible to industry, we propose a simple three-step process.
A framework is established that assesses the impact of variations in artificial intelligence (AI) data center (DC) loads on the fatigue damage of steam/gas turbines of the synchronous generators (SGs) from torsional oscillations. Next, a simple three-step process that is supported by frequency-domain analysis is laid out to quantify the limits on fluctuations in AI DC loads. In the first step, the maximum allowable variation in electrical power output at each SG terminal is independently determined from the first principles. This step needs only a lumped multi-mass model of the mechanical side of the SG. In the second step, we propose a new approach that relies on load flow to determine the so-called algebraic `interaction factor' that maps the change in AI DC load at a given bus to the corresponding change in each of the SG power outputs. In the third step, we propose a screening method to rank the candidate buses to site AI DCs and solve an optimization problem to determine the optimal allowable fluctuations in the AI DCs. We demonstrate the applicability of the proposed approach through frequency-domain and time-domain analyses in the modified IEEE 4-machine and IEEE-68 bus systems using a dynamic phasor framework. Finally, we demonstrate the scalability of the proposed approach on the synthetic 2000-bus Texas system.  

\end{abstract}


\section{Introduction}\label{sec:Intro} Artificial Intelligence (AI)  data centers (DCs) consume highly fluctuating power during their training cycles. At the beginning of a training cycle, the power consumption can increase by up to $70$ \% of the power rating in a short time followed by such fluctuations. Such large transients followed by these persistent variations in load consumption can pose challenges to the system reliability as highlighted by the North American Electric Reliability Corporation (NERC) \cite{NERC_2025}. 

Among different risks associated with the fluctuations, synchronous generator (SG) thermal turbine fatigue life reduction has been identified as one of the major issues. A review of recent papers shows that most of them have focused on other aspects of forced oscillations from AI DCs including excitation of electromechanical modes \cite{Biswas2025LargeLoadOscillations,Valverde2025ForcedOscillationsAI} and inverter-based resource (IBR)-induced subsynchronous oscillation (SSO) modes \cite{HossainChaudhuriLagoa2026DPSSO}. None of these works considered the turbine-generator fatigue life aspect, which is the focus of this paper. 

The fatigue life of the SG turbine primarily depends on two factors \cite{kundur1994power}: (A) Torsional oscillations of the turbine shafts and (B) Turbine blade vibrations due to off-frequency operations.  

\vspace{-10pt}
\subsection{Literature on Torsional Oscillations}\label{sec:TorsionLit} Torsional oscillations of SG turbines with the power grid have been extensively reported in the literature and can be divided into three categories as described below.

\textit{(a) Interaction with power system controls:} In $1969$, torsional mode destabilization was first observed at the Lambton generating plant in Ontario, Canada, due to control interaction with the power system stabilizer (PSS) \cite{WatsonCoultes1973,LeeKundur1986}. %The PSS was designed to damp a local mode of $1.67$ Hz frequency, but it produced a negative damping torque for the $16$ Hz torsional mode of the SG turbine. 
Other examples include instability through excitation control caused by terminal voltage limiter \cite{LeeKundur1986}, interaction with the speed governing system observed in $1983$ in Ontario Hydro \cite{Lee1985GovernorTorsionals}, and interaction with nearby HVdc rectifier converter controls 
% in Square Butte system in North Dakota destabilized a $11.5$ Hz torsional mode in the adjacent Milton Young generating station 
\cite{Bahrman1980HVDC}.  

\textit{(b) Subsynchronous resonance (SSR) with series compensated lines:} In the presence of a series-compensated line in electrical proximity of an SG, if the complement of the natural frequency (i.e., the synchronous frequency minus the natural frequency) of the network is close to one of the torsional frequencies of the turbine-generator mechanical system, then the torsional oscillations can be excited. This phenomenon called subsynchronous resonance led to the failure of the shaft system of the Mohave plant in Southern California, consecutively in $1970$ and $1971$ \cite{Concordia1973Subsynchronous,IEEECommittee1980SSR}. 

\textit{(c) Network switching:} Planned or unplanned network switching such as fault, fault clearing, line reclosing, and faulty synchronization can lead to excitation of torsional modes \cite{IEEEWG1980ScreeningGuide,IEEEWG1982SwitchingDisturbances}. In this context, steady state switching caused by delayed reclosing time of at least $10$ s has been categorized separately from successive network switching events (e.g., faults followed by clearance and high speed reclosing).  

\vspace{-10pt}
\subsection{Literature on Blade Vibrations}\label{sec:BladeVibLit} 
Each blade has many vibratory modes consisting of bending and torsion. In the presence of a flexible disk and identical blades (perfectly tuned), there are many more natural frequencies, many of them repeated. Resonant conditions are identified via nodal diameter map and  Campbell diagram. Blades of a turbine are excited by time varying forces because of a pressure field which is circumferentially non uniform. Assuming that this pressure field is time-invariant, each blade rotating with a constant angular speed experiences a periodic excitation with fundamental frequency equal to the rotational speed.  Therefore, excitation frequencies are not only the rotational speed , but also its integral multiples. In practice, blades are mistuned, which refers to small blade-to-blade variations due to manufacturing tolerances \cite{sinha2017vibration}. Because of mistuning, there can be a significant amplification of a blade’s vibratory amplitude. There are further complications when the angular speed is not a constant as it results in time-varying mistuning.  Recently, Sinha \cite{Sinha2026MistunedBladedDisk} has analysed the impact of fluctuations in rotor speed on the stability and forced response of a bladed disk. 

% The stimulus provided by natural vibrations in steam flow leads to a natural vibratory response of the turbine blade structural system. The blade vibration characteristics is composed of three modes -- tangential, axial, and torsional, which can be obtained from a Campbell diagram. The blade vibration modes lie in the supersynchronous range. The most critical blade sections are typically the last three rows of the low pressure (LP) turbine and occasionally the last row of the intermediate pressure (IP) turbine \cite{Akers1968Operation}.
\vspace{-10pt}
\subsection{Standard Remedies}\label{sec:Remedies}
Torsional oscillations cause cyclic variations in shear stress in turbine shafts. Negatively damped oscillations can lead to catastrophic shaft failure, whereas poorly-damped oscillations of large amplitude can cause reduction in turbine fatigue life. On the other hand, off-nominal frequency of operation may significantly increases the vibration amplitude in a nonlinear fashion, thereby reducing the fatigue life.

Over the years, standard guidelines have been developed in industry to address such issues \cite{kundur1994power}. In the context of torsional oscillations, standard tools have been developed to identify potential control interactions and SSR in the planning stage and numerous control design solutions \cite{AgrawalFarmer1979,UndrillKostyniak1976,EPRI2614} and power electronic hardware-based solutions (e.g., NGH dampers) \cite{Hingorani1981} have been successfully used.  

An IEEE working group report recommends that if a steady-state network switching event causes a sudden change of real power output from a generator below $50$ \% of its MVA rating, then it can be considered safe for turbine fatigue life \cite{IEEEWG1980ScreeningGuide}. More recommendations on successive network switching were developed in a separate IEEE report \cite{IEEEWG1982SwitchingDisturbances}.

Off-frequency operation is usually time-restricted based on blade design. Reference \cite{IEEE:C37.106-2003} uses composite characteristics of the safe frequency range of operation and the allowable time of safe operation based on five manufacturers. 

\vspace{-10pt}
\subsection{Recent Proposals from Utility Industry for Limiting AI DC Load Fluctuations}\label{sec:UtilityProposals} Persistent fluctuations in AI DC loads have the potential to excite torsional modes and cause off-nominal frequency operation that can potentially result in reduced turbine-generator fatigue life. Recently, the utility sector has come up with different regulations to limit these fluctuations. Some examples are mentioned below.

\noindent 1. Electric Reliability Council of Texas (ERCOT) \cite{ZhangRose2026LargeLoadPowerVariation}: Proposed that the large load power should not repetitively exceed $10$ MW change in a sliding time window of $5$ seconds.\\ 
2. Long Island Power Authority (LIPA) \cite{LIPA_LargeLoad_PerformanceRequirements}: Proposes limits the time-series output of the average zero-peak active power consumption by the large load in MW (P0-pk,avg), for the sum of any two adjacent frequency bins in the $5$ - $55$ Hz frequency range to $3.5$ MW for a rolling window $w$ of $10$ s. A frequency bin is the time-series output from the Fast Fourier Transform (FFT) on the active power with a $1$-s window. \\
3. Alberta Electric System Operator (AESO) \cite{ESIG2026LargeLoadPerformance}: The draft proposal requires that the variability in the forced oscillation should be limited to less than $16$ kW/$100$ ms, with a total permissible change limited to $160$ kW.\\
4. American Transmission Company (ATC) \cite{ESIG2026LargeLoadPerformance}: Proposes to limit active power oscillations for large loads over $200$ MW to $25$ MW over a period of less than $5$ s.\\
5. Southern Company (SOCO) \cite{SouthernCompany_TCLL}: Requires that loads minimize signal injection in the inter-area ($0.1$–$0.5$ Hz), sub-regional ($0.5$–$0.8$ Hz), local ($0.8$–$2$ Hz), and sub-synchronous ($5$–$55$ Hz) frequency bands, with automatic disconnection as a remedy if oscillations persist.

% \subsection{Literature on Forced Oscillations of AI DC Loads}\label{sec:AI_DC_Forced} 
% \begin{remark}
  
% \end{remark}

Although the general consensus is that the forced oscillations from large loads should not be allowed to cause amplification of subsynchronous oscillatory modes, the international standards highlight that the understanding of high frequency cycling is still significantly less mature compared to, say, voltage ride through. The gaps in this context are as follows:\\
1. The utility industry has not presented any solid justification behind the quantitative limits of the allowable variation in the power consumed by large loads.\\
2. Organizations like Energy Systems Integration Group (ESIG) have put forth two types of recommendations: (a) a study-based approach and (b) an approximation-based approach \cite{ESIG2026LargeLoadPerformance}. However, no technical method has been proposed for either. \\
3. To our knowledge, there is no method that optimizes the allowable fluctuations in AI DCs when multiple such AI DCs are considered in the context of turbine-generator fatigue life. This is relevant at the planning stage for ruling out buses with highly restrictive limits and also for imposing limits for existing AI DCs.

\vspace{-10pt}
\subsection{Contributions of this Paper}\label{sec:Contributions} To address these gaps, we propose a first principles-based approach to determine the allowable data center load variations. The following are the contributions of this work.\\
1. A framework is established to assesses the impact of variations in AI DC loads on the fatigue damage of the SGs from torsional oscillations.\\
2. A frequency-domain approach is proposed to determine the maximum allowable variation in electrical power at the generator terminal that does not endanger shaft life. This step only requires a generator's lumped multi-mass model and related data.\\
3. A simple approach based on load flow is proposed to calculate the so-called algebraic interaction factors (IFs) that determine how the change in a particular AI DC power consumption maps to the change of electrical power output at each generator terminal. \\
4. A screening method is proposed to select a group of candidate buses for siting the AI DCs and formulate an optimization problem, which can be solved to determine the optimal allowable fluctuations in a set of AI DCs in a system. This step involves a simple Linear Programming (LP) that is easy to implement and highly scalable.\\ % aids the siting of AI DCs along with imposing such limits on existing AI DCs. 
5. A dynamic phasor (DP)-based modeling framework is presented to validate the applicability of the proposed approach through frequency- and time-domain analyses in modified IEEE 4-machine and modified IEEE-68 bus systems \cite{Ameli-25-TPWRS}. Finally, we demonstrate the scalability of the proposed approach on the synthetic 2000-bus ERCOT system \cite{TAMU_ACTIVSg2000}.

\section{Proposed Framework to Assess Impact of AI DC Loads on Fatigue Damage  from Torsional Oscillations}\label{sec:Framework}
\begin{figure}
    \centering
    
    \begin{subfigure}[c]{0.24\textwidth}
        \centering
        \includegraphics[width=\textwidth]{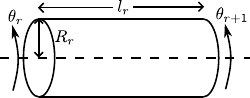}
        \caption{}
        \label{fig:Shaft}
    \end{subfigure}
    \hfill
    \begin{subfigure}[c]{0.24\textwidth}
        \centering
        \includegraphics[width=\textwidth]{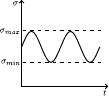}
        \caption{}
        \label{fig:Stress_variation}
    \end{subfigure}
    
    \vspace{-5pt}
    \caption{\small{(a) The $r$th shaft section; (b) A single-frequency stress variation with time.}}
    \label{fig:shaft_stress}
    \vspace{-10pt}
\end{figure}

In order to assess the impact of the fluctuations in AI DC loads that lead to corresponding power fluctuations at the SG terminals on turbine fatigue damage, 
a detailed multi-mass turbine has to be modeled for SG. There are multiple stages, such as high pressure (HP), intermediate pressure (IP), and low pressure (LP) stages in the turbine that are coupled to the generator rotor and the exciter. Each of these is connected by a varying size of shaft sections. The dynamics of the shaft system are defined by four sets of parameters: inertia constant $H$ of the individual masses, torsional stiffness $K$ of the shaft sections connecting adjacent masses, the self damping coefficient $\bar D$, and the mutual damping coefficient $D$. For the  $r$th shaft of length $l_r$ and circular cross-section with  radius $R_r$ shown in Fig. \ref{fig:Shaft} that connects the $r$th mass and the ($r + 1$)th mass, the torsional stiffness $K_r$ is equal to $\frac{GJ_r}{l_r}$, where $G$ is the shear modulus of elasticity of the shaft material (e.g., $G = 83\times10^9~Pa$ for steel) and $J_r = \frac{\pi R_r^4}{2}$ is the cross-sectional area moment of inertia.

Since the inertia constants of the masses are not identical to each other and the shafts connecting the masses are not perfectly rigid, the masses can oscillate relative to each other. The frequencies of these oscillations are associated with the torsional modes. There are multiple sources of damping of these modes, such as steam forces on turbine blades, shaft material hysteresis, energy dissipation in rotor bearings,  electrical damping, among others. 

A \emph{continuum model} of the rotor assembly can capture both subsynchronous and supersynchronous torsional vibration modes. However, as mentioned in \cite{kundur1994power}, the problems due to the interaction of the electrical and mechanical sides of the rotor are primarily restricted to the subsynchronous range, which can be captured using the \emph{lumped mass model}. In such models, the inertia constant of each rotor mass includes its share of shaft inertia, and it is assumed that turbine bladed disks are rigidly connected to the rotor shaft.  %Dynamic phasor (DP) framework is adopted in this work to model multi-mass turbine and further validation of the detailed approach.

\textit{Lumped mass model:} Let us assume that there are $N$ shaft sections connecting $N$+$1$ masses. %The $r$th shaft section connects the $r$th mass and the ($r$+$1$)th mass. 
The state-space representation of the $r$th mass  can be expressed using the following set of equations in per unit (p.u.)
% \small
% \begin{equation}
% \begin{aligned}
%     &\langle \dot{\delta}_k \rangle_0 = \langle \omega_k \rangle_0 - \omega_s \\
%     & \langle \dot{\omega}_k \rangle_0 = \frac{1}{2H_k}\{ \langle T_{mk} \rangle_0 - \langle T_e \rangle_0 - K_{k,k-1}(\langle \delta_k \rangle_0-\langle \delta_{k-1} \rangle_0)- K_{k,k+1}(\langle \delta_k \rangle_0-\langle \delta_{k+1} \rangle_0)-D_{k,k-1}(\langle \omega_k \rangle_0-\langle \omega_{k-1} \rangle_0)-D_{k,k+1}(\langle \omega_k \rangle_0-\langle \omega_{k+1} \rangle_0)-D_k \langle \omega_k \rangle_0\}
% \end{aligned}
% \end{equation}
% \normalsize
\small
\begin{equation}\label{eq:Lumped}
\resizebox{\columnwidth}{!}{$
\begin{aligned}
    & \dot{\delta}_r =  \omega_r - \omega_s \\
    & \dot{\omega}_r  =
    \frac{1}{2H_r}
    \left\{
    \begin{aligned}
        & T_{mr} - T_e - K_{r-1}\left( \delta_r - \delta_{r-1} \right)\\
        &- K_{r}\left( \delta_r - \delta_{r+1} \right)
        - D_{r-1}\left( \omega_r - \omega_{r-1} \right)\\
        &- D_{r}\left( \omega_r - \omega_{r+1} \right) 
        - \bar{D}_r \omega_r 
    \end{aligned}
    \right\}
\end{aligned}
$}
\end{equation}
\normalsize

\noindent where, $H_r$ and $\bar{D}_r$ are the inertia, and mechanical damping coefficient of the $r$th rotor mass, respectively; $D_{r}$ is the mutual mechanical damping coefficient for the $r$th section; $T_{mr}$ is the mechanical torque acting on mass $r$; $T_e$ is the electromagnetic torque; $\delta_r$ is the angular displacement of mass $r$ in a synchronously rotating frame in elect. rad.; $\omega_r$ is the angular speed of mass $r$ in elect. rad/s; and $\omega_s$ is the synchronous speed in elect. rad/s. Note that $T_e$ is present only in the generator mass, and $T_m$ is not present in both the generator and the exciter masses. %Please see the Appendix for more details on the basics of DP theory.

The SG with terminal voltage magnitude $V$ can be assumed to supply power to an infinite bus of voltage magnitude $E$ through a reactance $X$. To calculate the reactance $X$, the positive sequence bus impedance matrix $\bf{Z_{bus}}$ of the system (neglecting resistances) used for symmetrical short circuit analysis can be considered. The reactance value for the SG connected to the $i$th bus is $\bf{Z_{bus}}(i,i)$. The p.u. electromagnetic torque $ T_e $ is equal to the p.u. SG power output $P_e = \frac{EV}{X}~sin \delta_r $ and $r$ is $2$ for the mass of the generator rotor. In a small signal sense, this representation captures only the synchronizing component of $T_e$ to replicate the worst case scenario for the local mode. 

\begin{remark}
    The assumption here is that, for the SG connected to the grid, the controls including the PSS are designed in such a manner that it does not provide any negative contribution towards the damping of the torsional modes. Typically, this is standard practice.
\end{remark}

\textit{Maximum tensile stress calculation:} Assuming that the rotor shaft is in pure torsion, the maximum tensile  stress equals the maximum shear stress and is along the $45$ degree helical angle. Therefore, the maximum tensile stress, $\sigma_{r}$ in the $r$th shaft section %of length $l_r$ and circular cross-section with  radius $R_r$
shown in Fig. \ref{fig:Shaft}, 
% which connects the $r$th mass with the ($r$ + $1$)th mass, 
can be calculated using 
\begin{equation}\label{eq:Stress}
    \sigma_{r} = \frac{GR_r}{l_r}(\theta_{r+1}-\theta_{r})
\end{equation}
\noindent where, 
% $G$ is the shear modulus of elasticity of the shaft material (e.g., $G = 83\times10^9~Pa$ for steel), $J$ is the cross-sectional area moment of inertia, and $K_{eq}$ is the equivalent torsional stiffness. These quantities are related with each other through $K_{eq} = \frac{GJ}{l};~J = \frac{\pi r^4}{2}$. Let, $\theta_1$ and $\theta_2$ are the mechanical angle of rotation on both sides of the shaft. Note that 
$\theta_r$ (mech. rad) is equal to $\frac{2}{p_f}\delta_r$ and $p_f$ is the number of poles of the machine.

An example of the typical stress variation for single frequency SG power variation is shown in Fig. \ref{fig:Stress_variation}. The stress amplitude and mean can be calculated by
\begin{equation}
    \sigma_a = \frac{\sigma_{max}-\sigma_{min}}{2};~\sigma_m = \frac{\sigma_{max}+\sigma_{min}}{2}.
\end{equation}

\textit{Torsional fatigue characteristics:} It is well known that torsional modes are notoriously poor in damping with time constants ranging from $4$-$30$ s \cite{kundur1994power}. They vary with operating conditions and from one unit to another. Poorly-damped torsional oscillations cause large fluctuations in stresses and strains at a point within the shaft system. \textit{The process of progressive localized permanent structural change occurring in the shaft material subject to fluctuations that may culminate in cracks or complete fracture is known as fatigue}. 

\begin{figure}
	%\vspace{-5pt
	\centering
	\includegraphics[trim = {1.3cm 8cm 2.2cm 8cm}, clip,width= 0.4\textwidth]{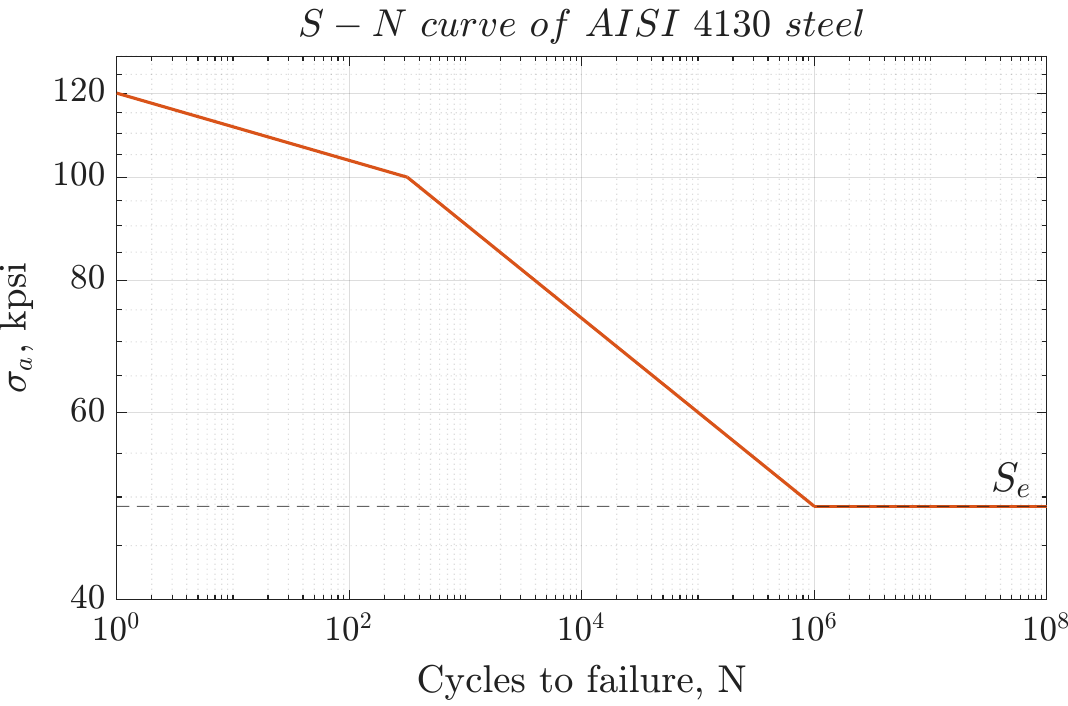}
	 \vspace{-5pt}
	\caption{\small{$S$-$N$ curve of AISI 4130 steel \cite{Norton2020MachineDesign}.}}
	\label{fig:S-N curve}
	\vspace{-15pt}
\end{figure}

A typical fatigue characteristic of a shaft specimen for AISI $4130$ steel, also called the $S$-$N$ curve or the W\"ohler curve, shown in Figure \ref{fig:S-N curve}, represents the relationship between the number of cycles to failure and the magnitude of fully reversed cyclic stress \cite{Norton2020MachineDesign}. The $Y$-axis represents the stress amplitude $\sigma_a$ when the mean stress $\sigma_m$ is zero. The figure also shows the high-cycle fatigue limit (HCFL) $S_e$, also known as the endurance limit of the material, below which practically no fatigue damage occurs. The goal is to keep $\sigma_a$ below $S_e$ so that no fatigue life is consumed. 

\begin{figure}
	%\vspace{-5pt
	\centering
	\includegraphics[width= 0.45\textwidth]{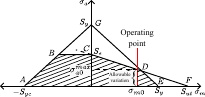}
	 \vspace{-5pt}
	\caption{\small{``Augmented'' modified Goodman diagram \cite{Norton2020MachineDesign}.}}
	\label{fig:Goodman}
	\vspace{-5pt}
\end{figure}

If the mean stress $\sigma_m$ is not zero, the maximum allowable stress amplitude will decrease following the ``augmented'' modified Goodman diagram \cite{Norton2020MachineDesign} as shown in Fig. \ref{fig:Goodman}. The diagram is used for fatigue and yielding assessment under combined mean stress $\sigma_m$ and alternating stress amplitude $\sigma_a$. The horizontal axis represents mean stress, spanning compressive (-ve) to tensile (+ve) values, while the vertical axis represents alternating stress. The triangular envelope $GAE$ defined by the yield stress $S_y$ describes static yielding in tension and compression, whereas the slanted line $CF$ connecting the endurance limit $S_e$ to the ultimate tensile strength $S_{ut}$ define the fatigue failure boundary for tensile mean stress (Goodman line). Line $BC$ defines fatigue failure in combination with compressive mean stress. The shaded region  inside these boundaries in Fig.~\ref{fig:Goodman} correspond to the safe operation, while points outside indicate yielding or fatigue failure depending on whether the failure line is governed by yield or fatigue criteria. Overall, the diagram combines yielding and fatigue limits to evaluate the safety of components subjected to cyclic loading with nonzero mean stress.

\textit{The proposed framework reveals whether a given fluctuation in $P_e$ (equivalently $T_e$ in p.u.) will lead to stress amplitudes within the safe limits of individual shaft sections.} We leverage this framework to propose an approach that helps determine limits on AI DC load fluctuations to avoid loss of turbine-generator fatigue life.    

\begin{figure}
	%\vspace{-5pt
	\centering
\includegraphics[width= 0.45\textwidth]{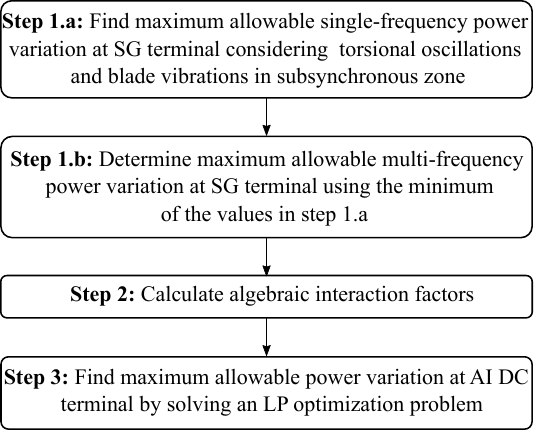}
	 % \vspace{-5pt}
	\caption{\small{Flowchart of the proposed approach for determining power variation limits of AI DC loads.}}
\label{fig:Proposed_approach}
	\vspace{-10pt}
\end{figure}

\section{Proposed Approach to Determine Limits on AI DC Load Fluctuations}\label{sec:Approach}
We propose a simple \textit{three-step} approach to determine the limit on the fluctuating AI DC loads so that the fatigue life of no SG in the system is negatively affected. The overall approach is summarized in Fig. \ref{fig:Proposed_approach}. Each step is elaborated in the following with greater detail.

\vspace{-10pt}
\subsection{Step I.a: Determine the Maximum Allowable Single-Frequency Power Variation at SG Terminal}\label{sec:StepI.a}
\subsubsection{Limits due to torsional oscillations}\label{sec:TorsionMono}
% \textit{Frequency-domain analysis of allowable power variation:} 
The transfer function $G_{r}(s)$ between the shear stress in the $r$th section $\sigma_{r}(s)$ and the power output $P_e(s)$ of the generator can be calculated from \eqref{eq:Stress} and the linearized model of \eqref{eq:Lumped}. Suppose that the following hold in the polar form in frequency domain:  $G_r(j\omega_i) = G_{ri}\angle \psi_{ri}$, $P_e(j\omega_i) = P_{ei}\angle \phi_i$, and $\sigma_r(j\omega_i) = \sigma_{ri}\angle (\psi_{ri} + \phi_i)$, respectively, where $G_{ri}$, $P_{ei}$, and $\sigma_{ri}$ are nonnegative real numbers and $0<\omega_i<\omega_s$. 
% The corresponding shear stress in the $r$th section is given by $\sigma_r(j\omega_i) = \sigma_{ri}\angle (\psi_{ri} + \phi_i)$.

Let, $\sigma_{ra0}^{max}$ be the maximum allowable amplitude of shear stress for the $r$th shaft section corresponding to the nominal steady state mean stress $\sigma_{rm0}$ as obtained from the Goodman diagram in Fig.~\ref{fig:Goodman}. Hence, the maximum allowable variation (amplitude) in $P_e$ for a single frequency ($\omega_i$) sinusoidal component considering torsional fatigue in the $r$th section is given by
\begin{equation}
    P_{ei,tor}^{max(r)} = \frac{\sigma_{ra0}^{max}}{G_{ri}}, ~~r = 1, 2, \cdots, N.
\end{equation}
Therefore, the maximum allowable variation in $P_e$ considering
each of the $N$ sections is given by 
\begin{equation}
    P_{ei,tor}^{max} = \min\limits_r P_{ei,tor}^{max(r)}, ~~r = 1, 2, \cdots, N.
\end{equation}

\subsubsection{Limits due to blade vibrations}
As mentioned in Sections \ref{sec:BladeVibLit} and \ref{sec:Remedies}, operation at off-nominal frequency causes vibrational stress in SG rotor blades. Hence, it is recommended to maintain the generator frequency deviation within a certain limit $\Delta f^{max}$ to avoid vibrational fatigue \cite{IEEE:C37.106-2003}. In this paper $\Delta f^{max} = 1.5$ Hz is chosen on the basis of \cite{IEEE:C37.106-2003}. If the transfer function between frequency variation, $\Delta f$ and generator electrical power output, $P_e$ is given by $G_f(s)$, then the maximum allowable single frequency ($\omega_i$) sinusoidal variation in $P_e$ for considering vibrational fatigue will be

\begin{equation}
    |P_{e,vib} (j\omega_i)|^{max} = \frac{\Delta f^{max}}{|G_f(j\omega_i)|} \triangleq P_{ei,vib}^{max}.
\end{equation}

Taking into account both torsional and vibrational fatigue, the maximum allowable single frequency ($\omega_i$) variation in $P_e$ will be 
\begin{equation}
    P_{ei}^{max} = \min \{P_{ei,tor}^{max}, ~P_{ei,vib}^{max}\}.
\end{equation}

% \vspace{-5pt}
\subsection{Step 1.b: Determine the Maximum Allowable Multi-Frequency Power Variation at SG Terminal}
% We can express the relationship between shaft stress at $r^{th}$ section, $\sigma_r(j\omega)$ and generator electrical torque, $T_e(j\omega)$ in frequency domain during steady state condition using transfer function, $G_r(j\omega)$. 
 In the presence of multi-frequency sinusoidal variations, the amplitude of the power output of the SG satisfies the following inequality
\begin{equation*}
\max\limits_t \left|\sum_i P_{ei} \sin(\omega_i t+\phi_i)\right| \leq \sum_i P_{ei}.
\end{equation*}
A sufficient condition to ensure that the amplitude of shear stress for any $r^{th}$ section in the presence of such variations in $P_e$ does not exceed $\sigma_{ra0}^{max}$ is given by the following inequality
\begin{equation*}
%\resizebox{\columnwidth}{!}{$
\begin{aligned}
 &\max\limits_t |\sum_i \sigma_{ri} sin(\omega_i t+ \psi_{ri}+\phi_i)| \leq \sum_i \sigma_{ri}\\
    & = \sum_i P_{ei} G_{ri}\leq \sigma_{ra0}^{max},~~~r = 1,2,...,N.
\end{aligned}
%$}
\end{equation*}
% where $\sigma_{ra0}$ is the maximum allowable amplitude of shear stress obtained from modified Goodman diagram corresponding to $r^{th}$ section.
% A sufficient condition to guarantee the above for each of the $N$ shaft sections is
% \begin{equation*}
%     \sum_i P_{ei} G_{ri} \leq \sigma_{ra0}^{max},~~~r = 1,2,...,N,
%\end{equation*}
Equivalently, we need to ensure that
\begin{equation*}
    \sum_i P_{ei}\frac{\sigma_{ra0}^{max}}{P_{ei,tor}^{max(r)}} \leq \sigma_{ra0}^{max},~~~r = 1,2,...,N.
\end{equation*}
% where, $|P_{ei,tor}|_{max(r)}$ is the value of maximum allowable amplitude of $P_e$ at $\omega_i$  for $r^{th}$ shaft section of the generator. 
The above condition simplifies to
\begin{equation*}
    \sum_i \frac{P_{ei}}{P_{ei,tor}^{max(r)}} \leq 1,~~~ r = 1,2,...,N.
\end{equation*}
A sufficient condition to meet the above set of constraints is the following
\begin{equation*}
    \sum_i \frac{P_{ei}}{P_{ei,tor}^{max}} \leq 1.
\end{equation*}
% Therefore, the sufficient condition is $\sum\limits_i \frac{|P_{ei}|}{|P_{ei,tor}|_{max}} \leq 1$, where $|P_{ei,tor}|_{max}=\min\limits_r|P_{ei,tor}|_{max(r)}$. 
According to H\"older's inequality, if $||\bold{a}||_\infty||\bold{P_e}||_1 \leq 1$, then $\bold{a}^T \bold{P_e} \leq 1$, where $\bold{a} = \left[\frac{1}{P_{e1,tor}^{max}}~\cdots~\frac{1}{P_{en,tor}^{max}}\right]^T$ and $\bold{P_e} = [P_{e1}~....~P_{en}]^T$.

Therefore, the sufficient condition is $\sum\limits_i P_{ei} \leq \frac{1}{\max\limits_i \frac{1}{P_{ei,tor}^{max}}}$, or equivalently, 
\begin{equation}
    \sum\limits_i P_{ei} \leq \min\limits_i P_{ei,tor}^{max}.
\end{equation} 
% Hence, the maximum sum of frequency components in electrical power output is $\min\limits_i |P_{ei,tor}|_{max}$ considering only torsional fatigue. 
If we take into account both torsional and vibrational fatigue, then the limit on the sum of the amplitudes of the (nonzero) frequency components in electrical power output of the generator will be given by
\begin{equation}
   \sum\limits_i P_{ei} \leq \min\limits_i P_{ei}^{max}.
\end{equation}
% where, $|P_{ei}|_{max} = \min\{|P_{ei,tor}|_{max},~|P_{ei,vib}|_{max}\}$.
Suppose $P_e^{max} = \inf \left\{P_{ei}^{max}| ~0< \omega_i < \omega_s\right\}$, then this allowable bound is given by
\begin{equation}\label{eq:SumPe}
   \boxed{\sum\limits_i P_{ei} \leq P_e^{max}.}
\end{equation}

\begin{remark}\label{rem:2}
    Since the lumped mass model is valid for the subsynchronous range, the above analysis considers $\omega_i < \omega_s$. Under the assumption that fluctuations in AI DC power $P_{DC}$ in the supersynchronous range will be practically attenuated before arriving at the SG terminals, a bound on the sum of the amplitudes of the frequency components in $P_e$ of each SG to avoid loss of fatigue life is given by \eqref{eq:SumPe}.
\end{remark}

% \textcolor{red}{Comment on steady state vs transients}

% \textbf{Special case:} If $G_r(j\omega_i) = M_{ri}$, where $M_{ri}\in \mathbb{R}$, then $\max\limits_t |\sum_i |T_{ei}| sin(\omega_i t+\phi_i)| \leq \frac{\sigma_{ra0}}{M_r}$ is a sufficient condition to ensure $\max\limits_t |\sum_i |\sigma_{ri}| sin(\omega_i t+\psi_{ri}+\phi_i)| \leq \sigma_{ra0}$ where $M_r = \max\limits_i |M_{ri}|$ and $\psi_{ri}\in\{0,\pi\}$. 

% \emph{Proof:} 
% \begin{equation*}
%     \begin{aligned}
%         &\max\limits_t |\sum_i |\sigma_{ri}| sin(\omega_i t+\psi_{ri}+\phi_i)|\\
%         =&\max\limits_t |\sum_i M_{ri}|T_{ei}| sin(\omega_i t+\phi_i)|\\
%         \leq&\max\limits_t \sum_i |M_{ri}|T_{ei}| sin(\omega_i t+\phi_i)|\\
%         =&\max\limits_t \sum_i |M_{ri}|||T_{ei}| sin(\omega_i t+\phi_i)|\\
%         \leq&M_r\max\limits_t \sum_i ||T_{ei}| sin(\omega_i t+\phi_i)|\\
%         \leq& \sigma_{ra0}
%     \end{aligned}
% \end{equation*}

% Hence, maximum allowable p-p electrical torque is $2\min\limits_r \frac{\sigma_{ra0}}{M_r}~~~~~\forall r = 1,2,3,....N$.  

\vspace{-10pt}
\subsection{Step 2: Calculate Algebraic IFs}
Upon determining the allowable variation in the real power output of each generator, the next step is to calculate the sensitivity of the change in real power at each SG terminal with respect to the change in the real power of each AI DC load. Suppose that a perturbation $P_{L}^{(j)}$ in the real power of the $j^{th}$ load (without changing any other loads) causes a corresponding change $P_{e}^{(i)}$ in the real power output of the $i$th generator. Then, the algebraic IF between the $i$th generator and the $j^{th}$ load is defined as $IF_{ij} = \frac{P_{e}^{(i)}}{P_{L}^{(j)}}$. 

We propose a new method for estimating the IFs that is based on load flow analysis. It has the following attributes.
\begin{enumerate}
    \item It is feasible to apply this in large-scale systems without needing sophisticated software tools. 
    \item It does not depend on decoupled load flow assumptions. 
    \item It can take into account the effects of constant power and constant impedance loads.
\end{enumerate}
The proposed steps are described below.

\emph{Step 2.1:} Run the base case load flow while modeling the constant power loads as $PQ$ loads and representing the constant impedance loads as the shunt admittances in bus data.

\emph{Step 2.2:} Using the load flow solution in Step 2.1, compute the internal voltage of each SG behind the subtransient impedance $R_a + jX_d^{''}$, where $R_a$ is the armature resistance and $X_d^{''}$ is the $d$-axis subtransient reactance. Assume that the internal voltage magnitudes and angles remain constant. 

\emph{Step 2.3:} Include the internal buses of the SGs in the bus data and update the $\bf{Y_{bus}}$ matrix with the subtransient impedances. Designate all the internal buses of SGs as slack buses and the terminal buses of SGs as $PQ$ buses while the characterization of the non-SG buses remain unchanged. Initialize the load flow voltage profile using the base case load flow solution. 

\emph{Step 2.4;} Change the power consumption by the $j^{th}$ load by the amount $P_{L}^{(j)}$. Run the load flow and observe the change $P_{e}^{(i)}$ in the $i$th SG. Calculate $IF_{ij}$ as defined earlier.
% Ideally, the load flow should converge without any iteration since the initial guess is the actual solution. Observe the power output, $P_{gen,i}$ of $i^{th}$ generator.

% \emph{Step 5:} Change load power consumption at bus $j$ by the amount $\Delta P_{load,j}$, and notice the change in $P_{gen,i}$ after running load flow again.  

% \emph{Step 6:} The final step is to calculate interaction factor using

% \begin{equation}
%     IF_{ij} = \frac{\Delta P_{gen,i}}{\Delta P_{load,j}}
% \end{equation}
% where, $\Delta P_{gen,i}$ is the change in generator power output at bus $i$ due to the change in load power input, $\Delta P_{load,j}$ at bus $j$.

\textit{Explanation of the proposed method:} The internal bus voltage magnitudes of the SGs can not change instantaneously following a perturbation due to the constant flux linkage theorem. The internal voltage angle cannot change instantaneously due to the rotor inertia. Subtransient reactance is considered since we are interested in the changes immediately following the perturbation. One slack generator is assumed to supply all the transmission losses in a typical load flow study, but all the generators share the transmission losses in the actual system. This aspect is taken into account in the proposed approach.

\vspace{-10pt}
\subsection{Step 3: Determine the Optimal Allowable Fluctuations in Presence of Multiple AI DC Loads} 
When multiple AI DCs are present in the system and/or are planned to be interconnected, we account for the worst case scenario that all the DC power variations could sum up in a way that can be worse for the system. The objectives are twofold -- (1) to determine a set of buses as candidates for siting AI DCs excluding the ones with existing AI DCs, and (2) to maximize the fluctuations in the AI DC loads while respecting the maximum allowable fluctuations in generator power outputs. We propose a two-step process for this.

\textit{Step 3.1:} \textit{Bus screening and ranking to site AI DCs that are not in the system:} We propose to use the upper bound on the maximum allowable fluctuations in power of each AI DC loads to achieve this. The maximum allowable sum of frequency components of power variation at $j^{th}$ AI DC load is defined as  $P_{dc}^{max(j)} = \min \left\{ \min \limits_i {\frac{P_e^{max(i)}}{IF_{ij}}},~~  P_{dc}^{comp(j)} \right \}$, which considers the presence of \textit{only the $j^{th}$ AI DC load}. Note that $P_e^{max(i)}$ corresponds to $P_e^{max}$ of $i^{th}$ SG and $P_{dc}^{comp(j)}$ is the \textit{worst case sum of the amplitudes of the (nonzero) frequency components observed in the real power consumption of the $j^{th}$ AI DC during the compute period without restrictions imposed}. Based on a typical fluctuation of AI DC load during the training phase in Fig. 3.8 of the NERC white paper \cite{NERC_2025}, we consider $P_{dc}^{comp(j)}$ to be $25$ \% of the AI DC rating, to be on the conservative side. We propose that the candidate buses are screened and ranked from the largest to the smallest $P_{dc}^{max}$ where the highest ranking (most preferred) corresponds to the bus with the largest $P_{dc}^{max}$. Planners may disregard locations that allow $P_{dc}^{max(j)}$ below a certain threshold. 

\textit{Step 3.2:} \textit{Optimal allowable fluctuations in selected AI DC loads:}  Let there be $k$ DC loads (including existing and those obtained from Step 3.1) and $m$ relevant SGs in the grid. We propose an iterative  optimization framework that needs to solve the following LP in the $q$th iteration
\begin{equation}\label{eq:LP_1}
\text{maximize}~~ \sum_{j=1}^k \left(P_{dc}^{(j)}\right)^{(q)}
\end{equation}
subject to
\begin{equation*}
\begin{aligned}
&\sum_{j=1}^k IF_{ij}\left(P_{dc}^{(j)}\right)^{(q)} \leq P_{e}^{max(i)},~~~~i = 1,2,...,m\\
& \alpha^{(q)} P_{dc}^{max(j)}  \leq \left(P_{dc}^{(j)}\right)^{(q)} \leq P_{dc}^{max(j)},~~~~j = 1,2,...,k\\
& \alpha^{(q)} = \alpha^{(q - 1)} - \beta, ~~~~ \alpha^{(q)}\in [0, 1],~~~~\beta \in (0, 1).
\end{aligned}
\end{equation*}

Here, $x^{(q)}$ denotes the variable $x$ corresponding to iteration number $q$, $P_{dc}^{(j)}$ is the \textit{maximum allowable sum of (nonzero) frequency components} the $j^{th}$ DC load, respectively. The decision variable in the LP is $P_{dc}^{(j)}$. In the first iteration, $\alpha^{(1)} = 1$ is chosen, followed by progressive reduction in an iterative manner. The value of parameter $\beta$ is defined by the user. The iterations stop when we obtain a feasible solution. 

\begin{remark}\label{rem:3}
    The LP solution determines the \textit{optimum limit on the sum of the amplitudes of the (nonzero) frequency components of the $j^{th}$ AI DC load $P_{dc}^{(j)}$ in the subsynchronous range.} An FFT can be performed on a $10$ s window of $P_{dc}^{(j)}$ to capture the frequency components at a $0.1$ Hz resolution. The sampling frequency should be high enough to capture the subsynchronous frequency content.
\end{remark}

\begin{remark}\label{rem:4}
    The frequency-domain analysis gives conservative \textit{steady-state} bounds. In reality, the fluctuations introduced by the AI data centers during the compute period will lead to transient overshoots in shear stresses of the shaft sections. Nevertheless, such overshoots will be limited since the ramp rate of transition from idle to compute stage and vise-versa is constrained by frequency regulation requirements. Moreover, the conservatism introduced by the norm inequalities gives us a reasonable degree of confidence that the transient overshoots will remain within the desired limits. Indeed, this has been proven to be the case based on exhaustive time-domain simulations.    
\end{remark}

\begin{remark}
    If the LP solution gives highly restrictive limits on certain buses, then one may discard those and rerun the LP with the remaining candidate buses. This process can continue until the optimal values are above a user-defined threshold.
\end{remark}

% Suppose that in the first step, the optimal values of decision variables $P_{dc}^{(j)}$ with $j \in \mathcal{M} \subset \left\{1, 2, \cdots m\right\}$ are equal to  $P_{dc}^{min(j)}$. In the second step, we remove the set $\mathcal{M}$ of AI DCs from the candidate AI DCs and run the LP with the reduced set of AI DCs.

\section{Dynamic Phasor (DP)-based Modeling Framework for Time-domain Validation}\label{sec:DP_model}
In this section, we present a very brief outline of the proposed DP-based modeling framework for the validation of the proposed approach in Section~\ref{sec:Approach}. The time-invariance of this framework makes it scalable to simulate large stiff systems with transmission line dynamics, SG stator transients, multi-mass SG turbine models, and IBR models. It also has the capability to perform frequency-domain analysis upon linearization.

\vspace{-5pt}
\subsection{Fundamentals of DP}\label{sec:DP_Funda}
The generalized averaging theory proposed in \cite{Verghese-91-DP} states that a near-periodic (possibly complex) time-domain waveform $x(\tau)$ in the interval $\tau \in (t - T, t]$ can be expressed using a Fourier series of the form $x(\tau) = \sum_{k = -\infty}^{\infty} X_k(t) e^{jk\omega_s\tau}$, where $X_k(t)$ are the complex time-varying Fourier coefficients obatined from a moving window of width $T$ sliding over the signal, and $\omega_s = \frac{2\pi}{T}$, $k \in \mathbb{Z}$. The $k$th coefficient is called the \textit{$k$th DP}, which can be calculated in time $t$ by the following \textit{averaging} operation $X_k(t) = \frac{1}{T}\int_{t-T}^{t} x(\tau) e^{-jk\omega_s\tau}d\tau = \left \langle x \right \rangle_k (t)$. The DP framework provides a good approximation of the signal $x(\tau)$ using the set $\mathcal{U}$ of dominant Fourier coefficients such that $x(\tau) \approx  \sum_{k \in \mathcal{U}} \left \langle x \right \rangle_{k}(t) e^{jk\omega_s\tau}$. Therefore, the generalized averaging-based method leads to an approximated model. We remove the time variable from the DP notation for the sake of simplicity. The following are some of the useful properties of DPs:\\
 (1) $    \left \langle \frac{\mathrm{d} x}{\mathrm{d} t} \right \rangle_k = \frac{\mathrm{d} \left \langle x \right \rangle_{k}}{\mathrm{d} t} + jk\omega_s\left \langle x \right \rangle_{k}$.\\
 (2) if $x(\tau)$ is real, then $    \langle x \rangle_k=\langle x \rangle_{-k}^*$. In $pnz$ frame,  $\langle x_p \rangle_{-k} = \langle x_n \rangle_k^*$.

\begin{figure}
	%\vspace{-5pt
	\centering
	\includegraphics[width= 0.45\textwidth]{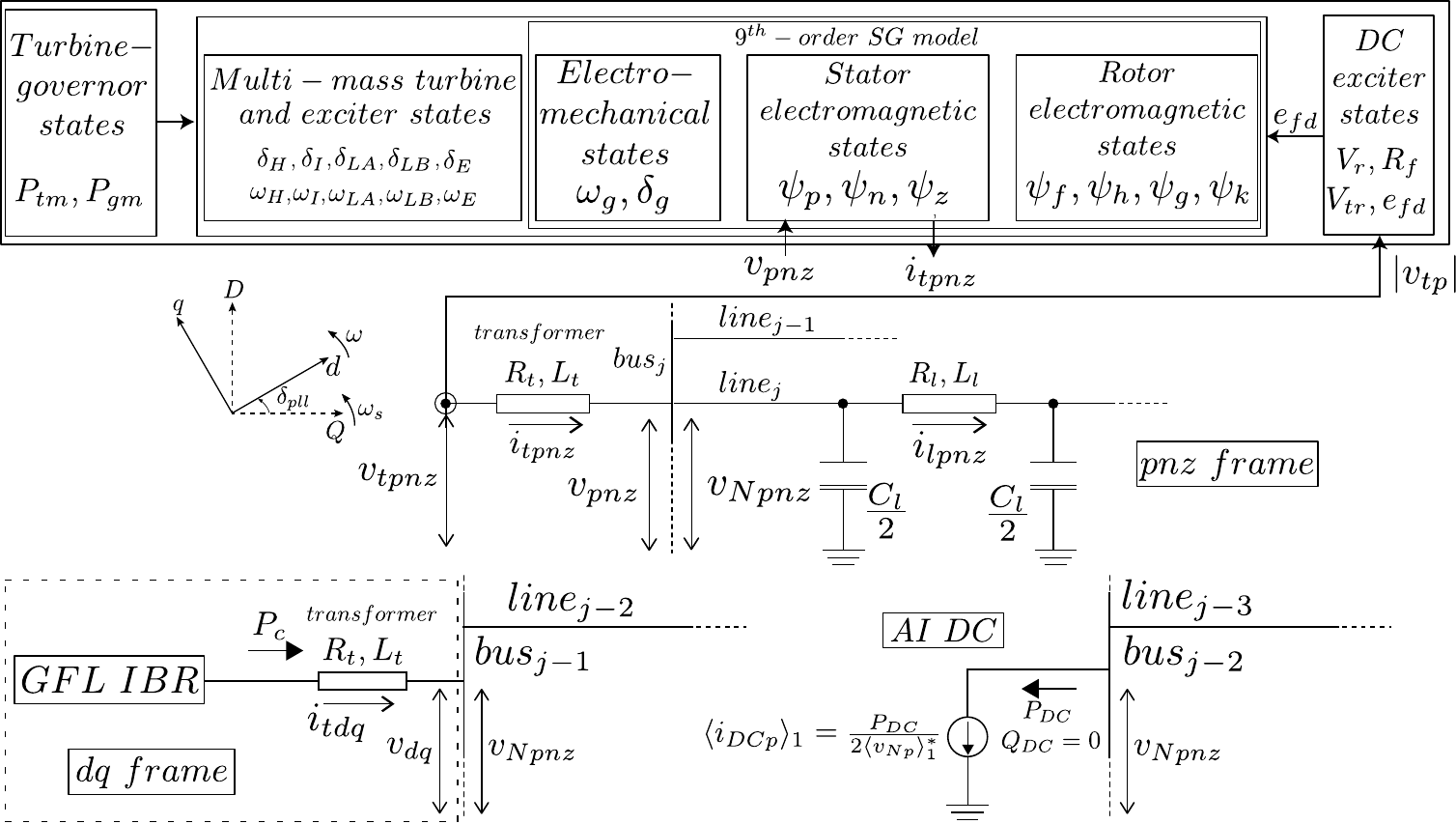}
	 \vspace{-5pt}
	\caption{\small{Generalized DP-based modeling framework for multi-mass SGs, transmission  system, AI DC loads, and GFL IBRs. %[parameters: $\tau_{c}$ = $0.05s$, , $C_{c}$ = $1.7370~pu$, $k_{dc}$ = $1080~pu$, $R$ = $0.0056~pu$, $L$ = $0.2454~pu$, $R_t$ = $0~pu$, $L_t$ = $0.15~pu$ $S_{base}$ = $100MVA$, $V_{dc,base}$ = $48.98kV$, $V_{ac,base}$ = $20kV$].
    }}
	\label{fig:DP_framework}
	\vspace{-10pt}
\end{figure}

The time-domain waveform $x(\tau)$ can be the $abc$ phase quantities, asynchronous $dq0$ frame quantities, or synchronous $DQ0$ frame quantities (note the orientation of the axes used in Fig.~\ref{fig:DP_framework}). 
% Using the transformation in synchronously rotating reference frame, the relationship between $DQ0$ and +ve/-ve/zero sequence ($pnz$) quantities can be obtained in terms of DPs. 
%\vspace{-3pt}
% \small
% \begin{equation}\label{eqn:DQtopnz}
% \resizebox{\columnwidth}{!}{$
%     \begin{aligned}
%         &\langle x_D\rangle_k=\frac{\langle x_p\rangle_{k+1}+\langle x_n\rangle_{k-1}}{\sqrt{2}};~
%         \langle x_Q\rangle_k=-\frac{\langle x_p\rangle_{k+1}-\langle x_n\rangle_{k-1}}{\sqrt{2}j}\\
%         &\langle x_p\rangle_{k+1}=\frac{\langle x_D\rangle_{k}-j\langle x_Q\rangle_{k}}{\sqrt{2}};~~
%         \langle x_n\rangle_{k-1}=\frac{\langle x_D\rangle_{k}+j\langle x_Q\rangle_{k}}{\sqrt{2}};~\langle x_z \rangle_k = \langle x_0 \rangle_k 
%     \end{aligned}
%     $}
% \end{equation}
\normalsize
%\vspace{-3pt}
% The following equations describe the relationship between asynchronous $dq$ frame quantities and $pnz$ frame quantities
% \vspace{-3pt}
% \small
% \begin{equation}\label{eqn:pnztodq}
% \begin{aligned}
%     &\langle x_d \rangle_k = \frac{1}{\sqrt{2}j}\left(e^{j\delta_{pll}}\langle x_n \rangle_{k-1}-e^{-j\delta_{pll}}\langle x_p \rangle_{k+1}\right)\\
%     &\langle x_q \rangle_k = \frac{1}{\sqrt{2}}\left(e^{j\delta_{pll}}\langle x_n \rangle_{k-1}+e^{-j\delta_{pll}}\langle x_p \rangle_{k+1}\right) \\
%     &\langle x_p \rangle_{k+1} = \frac{1}{\sqrt{2}}e^{j\delta_{pll}}\left(\langle x_q \rangle_k-j\langle x_d \rangle_k\right)\\
%     &\langle x_n \rangle_{k-1} = \frac{1}{\sqrt{2}}e^{-j\delta_{pll}}\left(\langle x_q \rangle_k+j\langle x_d \rangle_k\right)
% \end{aligned}
% \end{equation}
% \normalsize
% \vspace{-3pt}
% where, $\delta_{pll}$ is the angle between the $Q$-axis and the $d$-axis. %For further details on the properties of DPs please refer to [cite].

\vspace{-5pt}
\subsection{Generalized DP-based Modeling Framework}
%We propose a novel DP-based modeling approach to simulate multi-machine system involving IBRs. The IBRs are modeled in $dq$ frame, and the rest of the network including the SGs are modeled in $pnz$ frame. 
We use the DP framework as shown in Fig.~\ref{fig:DP_framework} to model the test system and validate the proposed approach. It has the following features (readers can refer to \cite{HossainChaudhuriLagoa2026DPSSO} for more details).

1. \emph{Modeling of the Network, Loads,  and SG in the $pnz$ domain:} Transmission lines, loads, and SGs are modeled in the $pnz$ domain, where $k=\pm1$ DPs are considered to capture the subsynchronous frequency regime. The mechanical side of SGs including multi-mass turbines is modeled using $k=0$ DP since the mechanical variables vary slowly. The DC loads are functionally modeled as +ve sequence constant power loads with unity power factor, see Fig.~\ref{fig:DP_framework}.

2. \emph{Modeling of Grid-following (GFL) IBR in the $dq$ frame:} The controllers of GFL IBRs are typically based on vector control in the $dq$ frame. 
%Hence, it is logical to consider the dominant DP coefficients in the same $dq$-frame for the averaged model of IBR circuit and the controllers instead of representing them in another frame. 
DP coefficients $k=0$ and $k=\pm2$ are chosen to capture positive and negative sequence quantities, respectively. The $dq$ frame associated with a GFL IBR is uniquely determined by its phase locked loop (PLL).  %The GFL IBR model is interfaced with the $pnz$ frame based network using \eqref{eqn:pnztodq}.

\section{Results and Discussions}\label{sec:Results}

% 1. EMT-DP model validation\\
% i. Eigenvalues comparison: Dr. Kulkarni paper\\
% ii. Time domain result simulink model (FBM paper variables)

% 2. Application on IEEE 4-machine 2-area system: step 1.a: step 1.b: step 2:(Minimum Pe using isolated model and minimum Pdc using detailed model and show their relationship using IF.)Interaction factor heatmap and comparison with bode plot  step 3: LP Validation: Time domain validation using Rainflow and Palmgren miner rule.

% 3. Repeat for IEEE-68 bus system

% 4. Texas synthetic system

\begin{figure}
	%\vspace{-5pt
	\centering
	\includegraphics[width= 0.4\textwidth]{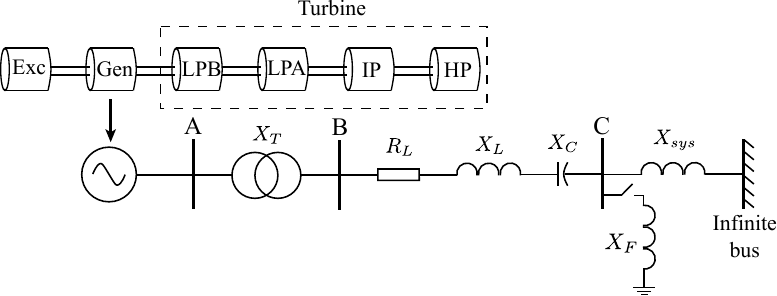}
	 \vspace{-5pt}
	\caption{\small{IEEE First benchmark model for SSR \cite{IEEESSRBenchmark1977}.}}
	\label{fig:IEEE FBM}
	\vspace{-10pt}
\end{figure}

% \begin{table}
% %\small
% \caption{Torsional and network modes for FBM}
% %\normalsize
% \label{tab:linearization}
% %\renewcommand{\arraystretch}{1.2}
% \centering
% \resizebox{0.35\textwidth}{!}{
% \begin{tabular}{|c|c|c|}
% \hline
% %\small
% \multicolumn{3}{|c|}{\textbf{Torsional modes}} \\ \hline
% \textbf{Mode} & \textbf{DP model} & \textbf{Benchmark}\\  \hline
% 1 & 0.02$\pm$j99.5 & 0.02$\pm$j99.5\\ \hline
% 2 & 0.008$\pm$j127.1 & 0.008$\pm$j127.1\\ \hline
% 3 & 0.59$\pm$j159.8 & 0.59$\pm$j159.8\\ \hline
% 4 & 0.004$\pm$j202.8 & 0.005$\pm$j202.8\\ \hline
% 5 & $\pm$j298.2 & $\pm$j298.2\\ \hline
% \multicolumn{3}{|c|}{\textbf{Network modes}} \\ \hline
% \textbf{Mode} & \textbf{DP model} & \textbf{Benchmark}\\  \hline
% 1 & -3.70$\pm$j155.65 & -3.70$\pm$j155.64\\ \hline
% 2 & -4.69$\pm$j155.38 & -4.71$\pm$j154.56\\ \hline
% 3 & -53.55$\pm$j230.85 & -53.58$\pm$j230.81\\ \hline
% 4 & -53.55$\pm$j523.15 & -53.58$\pm$j523.17\\ \hline
% 5 & -4.69$\pm$j598.83 & -4.69$\pm$j598.82\\ \hline
% 6 & -3.02$\pm$j598.62 & -3.09$\pm$j599.39\\ \hline

% \end{tabular}
% }
% %\normalsize
% \vspace{-5pt}
% \end{table}

\begin{table}[h]
\centering
\caption{Torsional and network modes for IEEE FBM}\label{tab:linearization}
\begin{tabular}{|c|c|c|c|}
\hline
\multicolumn{2}{|c|}{Torsional modes} & \multicolumn{2}{c|}{Network modes} \\
\hline
DP model & \cite{DP_SSR} & DP model & \cite{DP_SSR} \\
\hline
0.02$\pm$j99.5   & 0.02$\pm$j99.5   & -3.70$\pm$j155.65 & -3.70$\pm$j155.64 \\
0.08$\pm$j127.1  & 0.08$\pm$j127.1  & -4.69$\pm$j155.38 & -4.71$\pm$j154.56 \\
0.59$\pm$j159.8  & 0.59$\pm$j159.8  & -53.55$\pm$j230.85 & -53.58$\pm$j230.81 \\
0.004$\pm$j202.8 & 0.005$\pm$j202.8 & -53.55$\pm$j523.15 & -53.58$\pm$j523.17 \\
$\pm$j298.2       & $\pm$j298.2       & -4.69$\pm$j598.83 & -4.69$\pm$j598.82 \\
                  &                   & -3.02$\pm$j598.62 & -3.09$\pm$j599.39 \\
\hline
\end{tabular}
\end{table}

\begin{figure}
    \centering
    
    \begin{subfigure}[!t]{0.4\textwidth}
        \centering
        \includegraphics[trim={1.8cm 9cm 2cm 8.1cm}, clip, width=\textwidth]{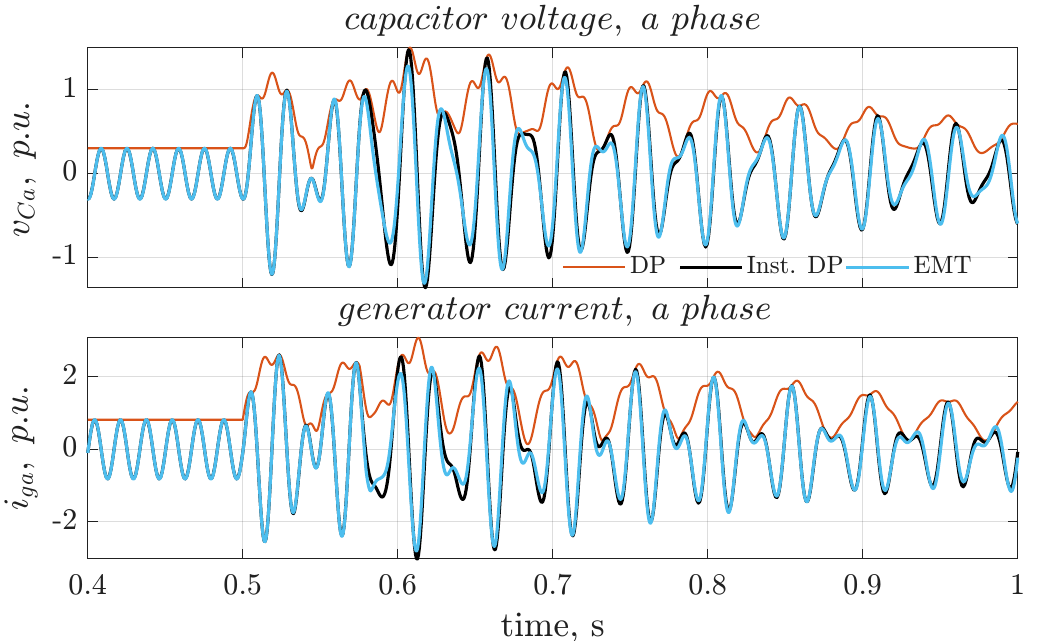}
        \vspace{-15pt}
        \caption{}
        \label{fig:dp_emt_validation_1}
    \end{subfigure}
    \hfill
    \begin{subfigure}{0.4\textwidth}
        \centering
        \includegraphics[trim={1.5cm 9cm 2cm 8.2cm}, clip, width=\textwidth]{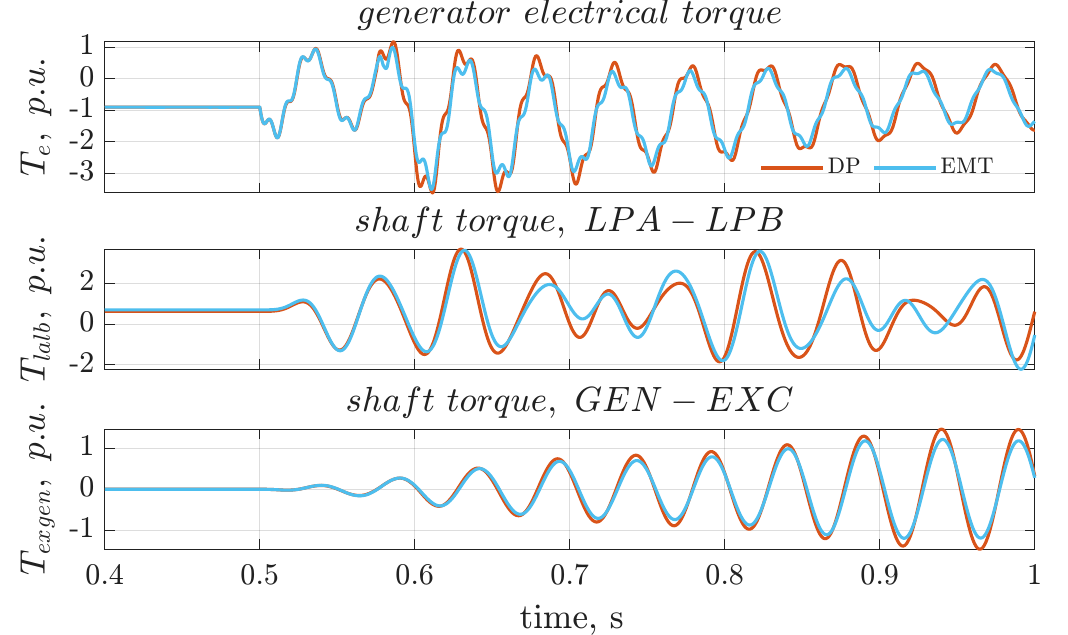}
        \vspace{-15pt}
        \caption{}
        \label{fig:dp_emt_validation_2}
    \end{subfigure}
    
    \vspace{-8pt}
    \caption{\small{Comparison of responses from DP model and EMT model of IEEE FBM. Inst. DP: $x(\tau) \approx  \sum_{k \in \mathcal{U}} \left \langle x \right \rangle_{k}(t) e^{jk\omega_s\tau}$. }}
    \label{fig:dp_emt_validation}
    \vspace{-7.5pt}
\end{figure}

\subsection{Validation of the DP Framework}\label{sec:DP_Valid}
To gain confidence in the accuracy of the modeling approach using the DP framework, we have performed both frequency- and time-domain analyses for the IEEE First benchmark model (FBM) for SSR \cite{IEEESSRBenchmark1977} as shown in Fig. \ref{fig:IEEE FBM}. The eigenvalues of the linearized DP model are shown in Table \ref{tab:linearization} and compared to the eigenvalues presented in \cite{DP_SSR}. Moreover, a three phase-to-ground (LLL-G) fault is applied at bus C in both the DP and a publicly available EMT model of the system in Matlab/Simscape \cite{Young2026SSR}. The simulation results are compared in Fig. \ref{fig:dp_emt_validation}, where a close match is observed between the responses.

% \begin{figure}
% 	%\vspace{-10pt}
% 	\centering
% 	\includegraphics[trim = {2cm 9cm 2cm 8cm}, clip,width= 0.45\textwidth]{Figures/DP_emt_validation_fig1.pdf}
% 	 \vspace{-10pt}
% 	\caption{Comparison of DP and EMT model of FBM.}
% 	\label{fig:dp emt validation}
% 	\vspace{-7.5pt}
% \end{figure}

\begin{figure}[!t]
	%\vspace{-5pt
	\centering
	\includegraphics[width= 0.4\textwidth]{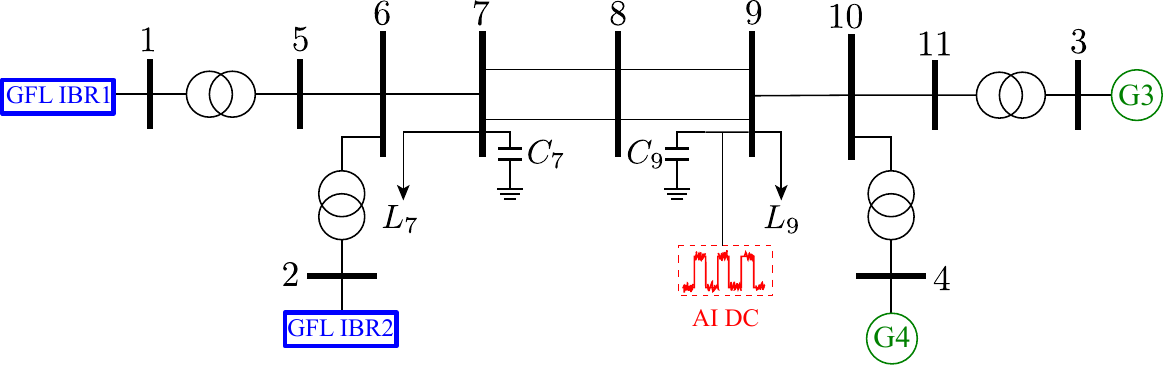}
	 \vspace{-5pt}
	\caption{\small{Modified IEEE 4-machine system \cite{hossain2025dynamicphasorframeworkanalysis}.}}
	\label{fig:IEEE 4-machine system}
	\vspace{-10pt}
\end{figure}

\begin{figure}[!t]
	%\vspace{-10pt}
	\centering
	\includegraphics[trim = {1.3cm 7cm 1.6cm 8cm}, clip,width= 0.4\textwidth]{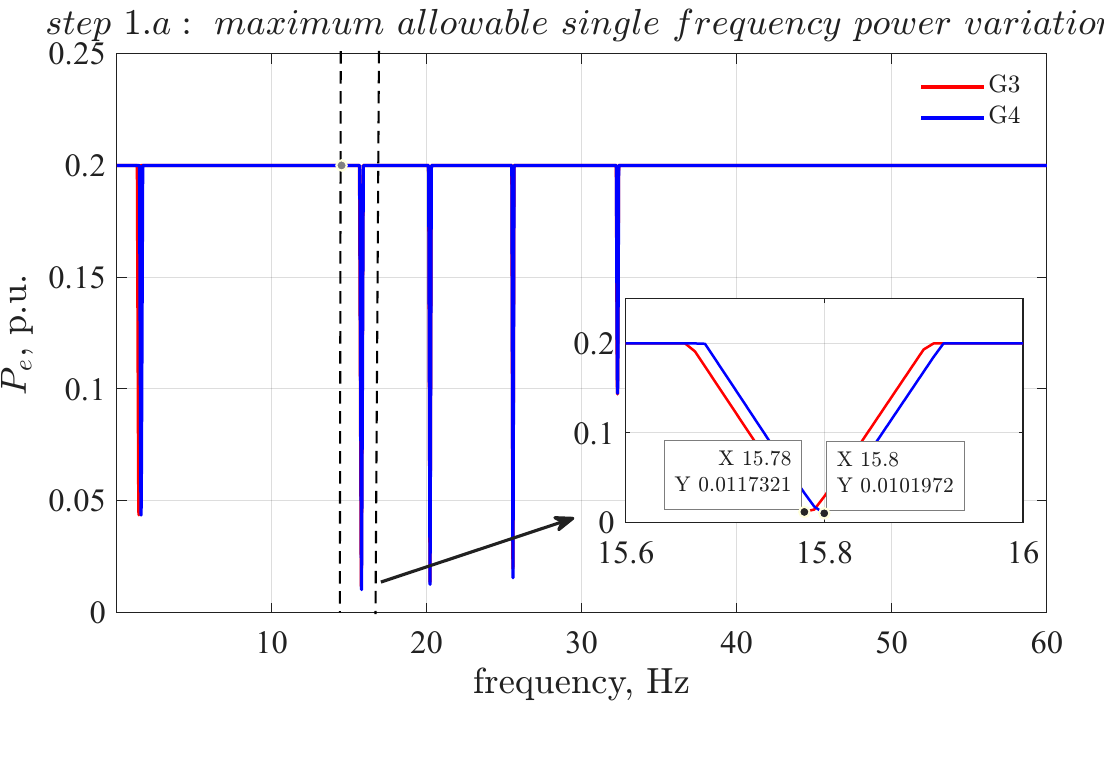}
	 \vspace{-10pt}
	\caption{\small{Step 1.a for modified IEEE 4-machine system (p.u. on SG base). Notches correspond to torsional mode frequencies.}}
	\label{fig:4mac_isolated}
	\vspace{-15pt}
\end{figure}

% \begin{figure}
% 	%\vspace{-10pt}
% 	\centering
% 	\includegraphics[trim = {1.8cm 7cm 2cm 7.5cm}, clip,width= 0.4\textwidth]{Figures/IF_bode_4mac.pdf}
% 	 \vspace{-10pt}
% 	\caption{Comparison of actual IF and algebraic IF.}
% 	\label{fig:comp_IF_4mac}
% 	\vspace{-7.5pt}
% \end{figure}

\begin{table}[h]
\centering
\caption{\small{$P_{dc}^{max(j)}$ vs $P_{dc,actual}^{max(j)}$ in MW}}
\label{tab:4mac_IF}
\resizebox{0.48\textwidth}{!}{
\begin{tabular}{|c|c|c|c|c|c|}
\hline
\textbf{DC Bus} & \textbf{SG} & \textbf{$P_e^{max(i)}$} & \textbf{Algebraic IF} & \textbf{$P_{dc}^{max(j)}$} & \textbf{$P_{dc,actual}^{max(j)}$} \\
\hline
9 & G3 & 10.56 & 0.1441 & \multirow{2}{*}{32.39} & \multirow{2}{*}{71.37} \\
\cline{1-4}
9 & G4 & 9.18 & 0.2834 &  &  \\
\hline
7 & G3 & 10.56 & 0.4875 & \multirow{2}{*}{11.47} & \multirow{2}{*}{74.46} \\
\cline{1-4}
7 & G4 & 9.18 & 0.8004 &  &  \\
\hline
\end{tabular}
}
\vspace{-5pt}
\end{table}

\begin{figure}[!t]
    \centering
    % First row
    \begin{subfigure}{0.24\textwidth}
        \centering
\includegraphics[trim={0.8cm 7cm 1cm 7cm}, clip, width=\textwidth]{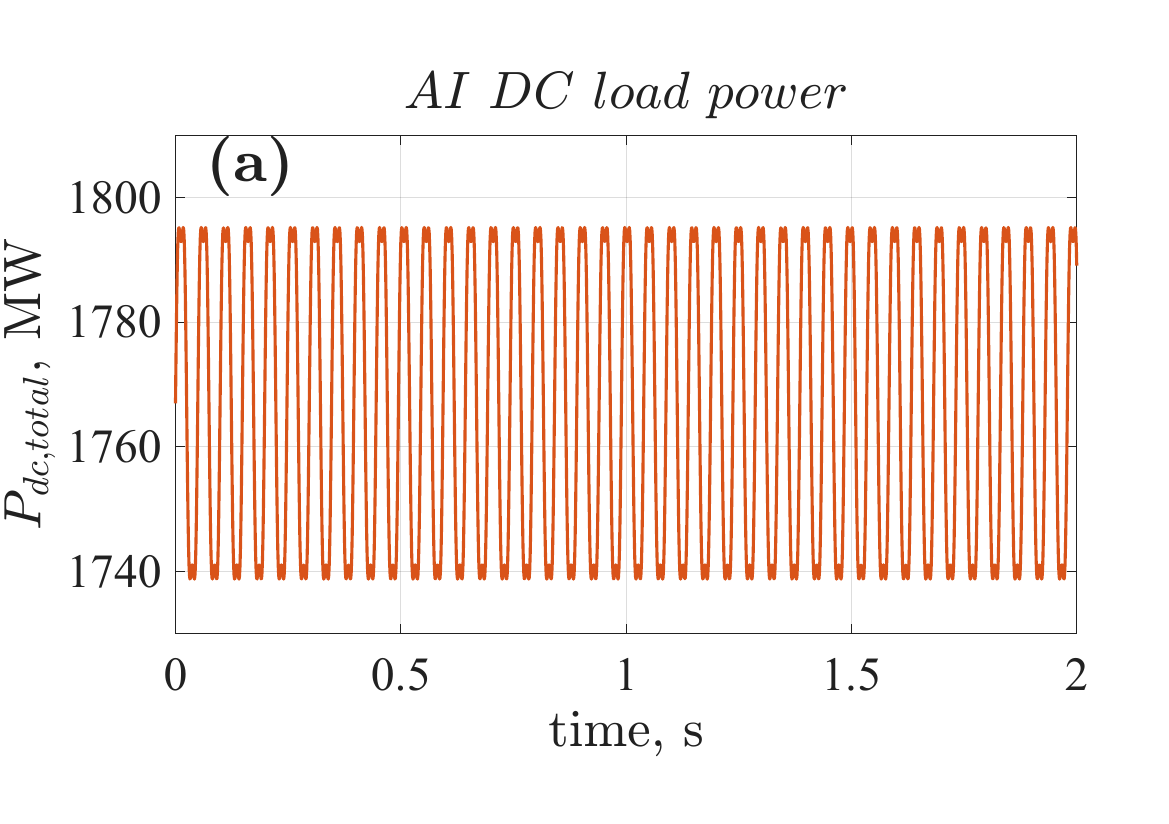}
        \vspace{-18pt}
        %\caption{}
    \label{fig:sub1}
    \end{subfigure}
    \hfill
    \begin{subfigure}{0.24\textwidth}
        \centering
\includegraphics[trim={1cm 7cm 1cm 7cm}, clip, width=\textwidth]{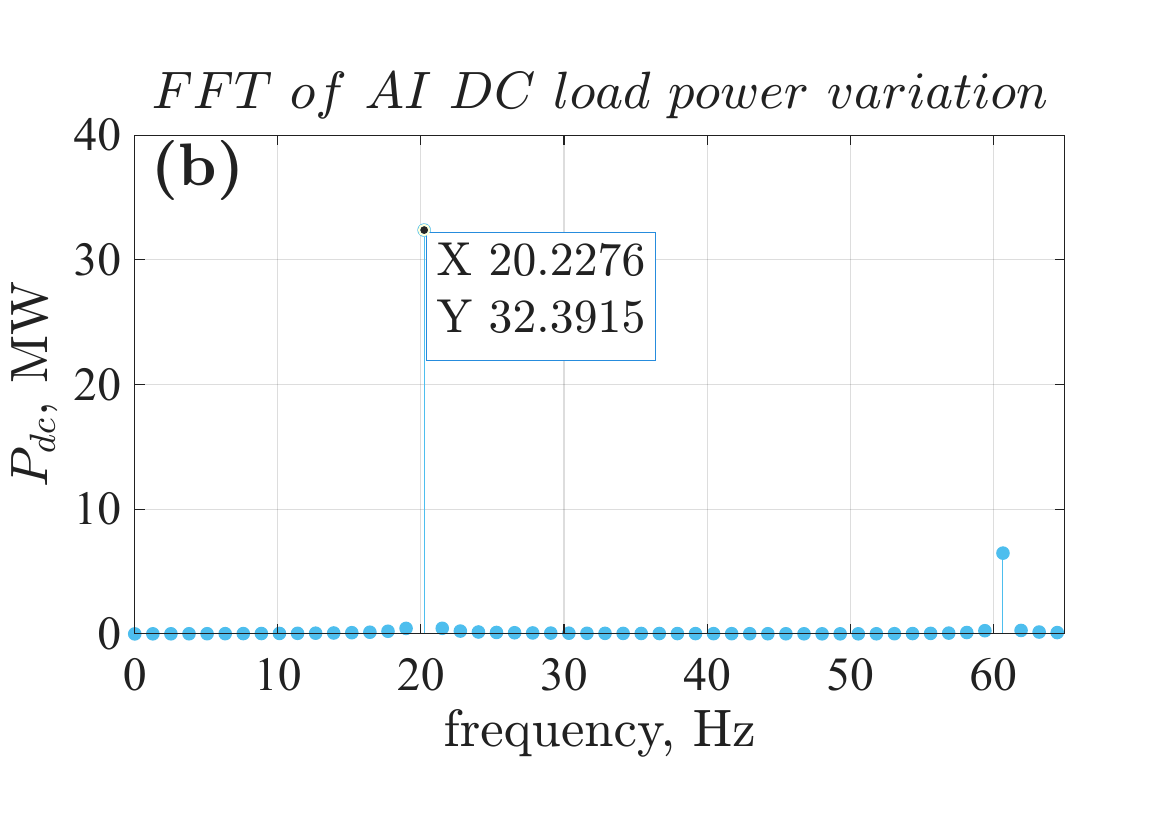}
        \vspace{-18pt}
        %\caption{}
\label{fig:sub2}
    \end{subfigure}
    
   % Second row
    \begin{subfigure}{0.24\textwidth}
        \centering
\includegraphics[trim={1cm 7cm 1cm 8cm}, clip, width=\textwidth]{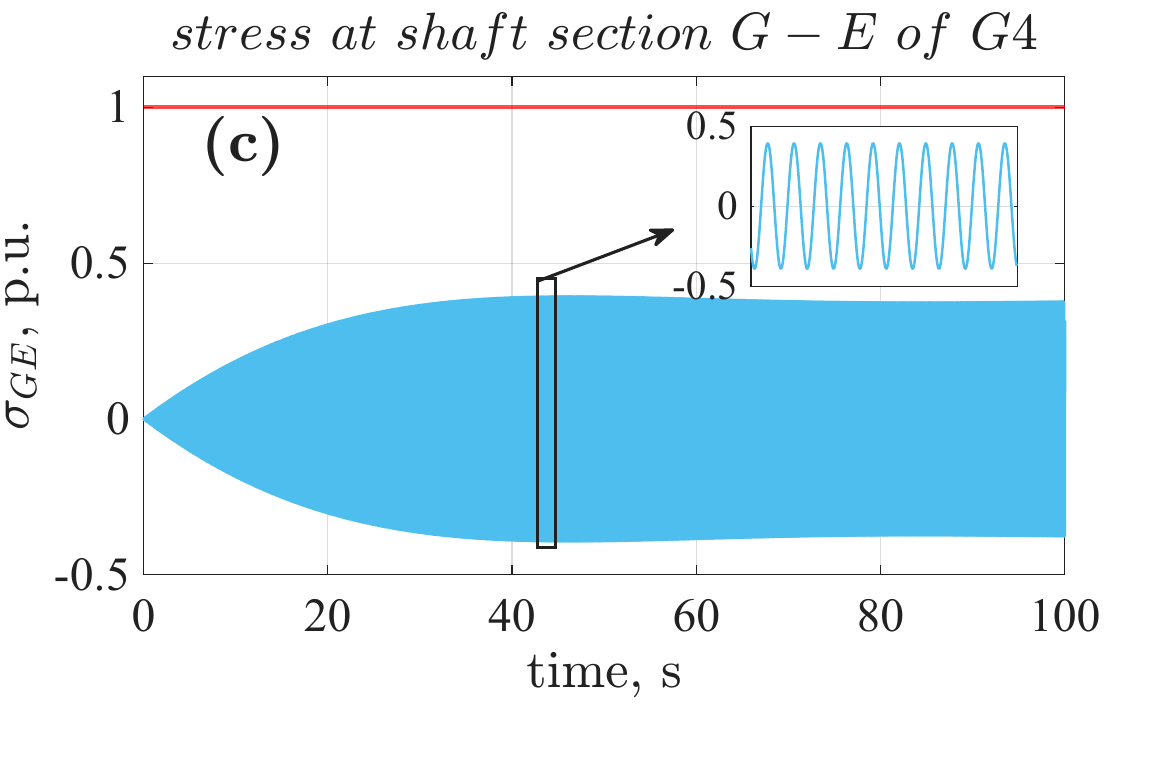}
        \vspace{-20pt}
        %\caption{}
    \label{fig:sub3}
    \end{subfigure}
    \hfill
    \begin{subfigure}{0.24\textwidth}
        \centering
\includegraphics[trim={1cm 7cm 1cm 8cm}, clip, width=\textwidth]{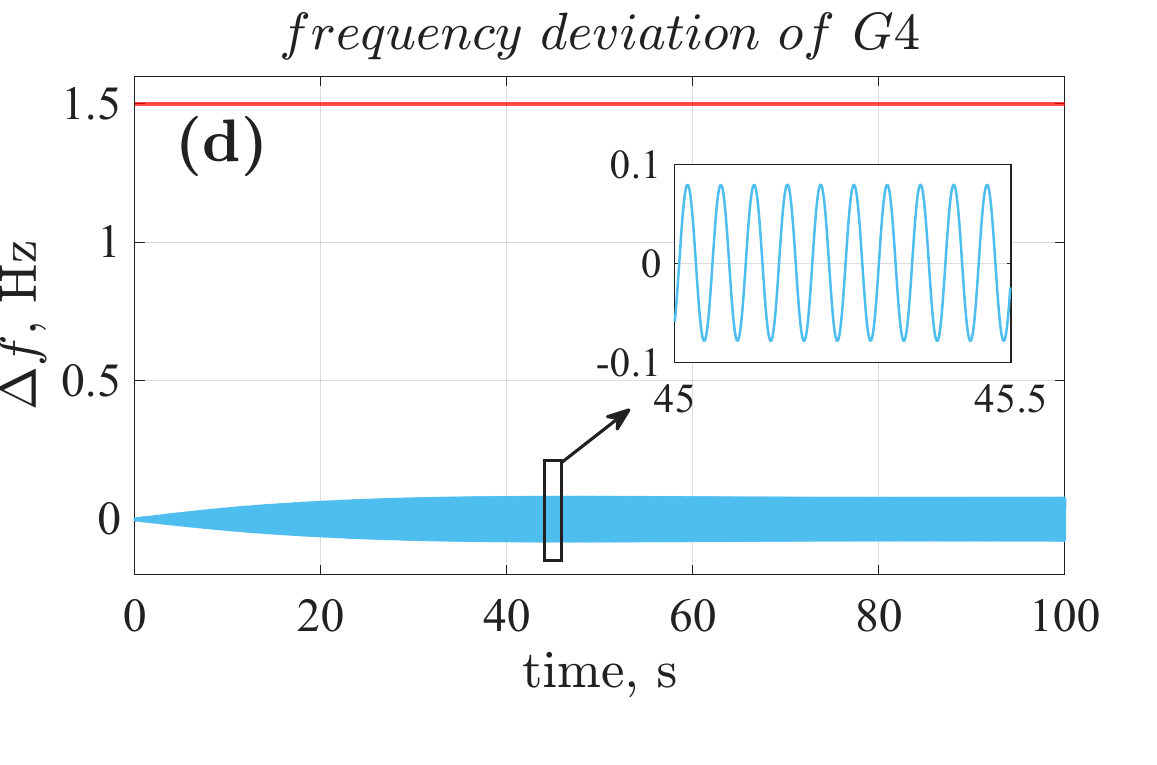}
        \vspace{-20pt}
        %\caption{}
        \label{fig:sub4}
    \end{subfigure}
    \caption{\small{(a) AI DC load power in MW in bus $9$; (b) FFT of (a); (c) p.u. stress (cyan) in generator-exciter (G-E) shaft section of G4 with the maximum allowable value (orange) obtained from Fig.~\ref{fig:Goodman} and taken as base, and (d) frequency deviation (cyan) of G4 overlapped with the maximum allowed (orange).}}
\label{fig:4mac_time_domain}
    \vspace{-7.5pt}
\end{figure}
\vspace{-10pt}
\subsection{Application on the Modified IEEE 4-machine System}\label{sec:4mac}
Figure \ref{fig:IEEE 4-machine system} shows the modified IEEE-4 machine system \cite{hossain2025dynamicphasorframeworkanalysis} where the SGs at buses $1$ and $2$ are replaced with two GFL IBRs of the corresponding ratings. The mechanical sides of $G3$ and $G4$ are modeled using parameters from \cite{IEEESSRBenchmark1977}. However, the damping coefficients were chosen so that the time constants of the torsional modes remain within $4$-$30~s$ \cite{kundur1994power}. 
\subsubsection{Step 1.a:  Maximum allowable single-frequency power variation at SG terminal} Figure \ref{fig:4mac_isolated} shows the maximum allowed electrical power variations at the SG terminals (p.u. on the SG base) to avoid loss of fatigue life, as a function of frequency. We have imposed an absolute upper limit of 20\% of the SG rating on this allowed variation, which can be defined by the user, and is inconsequential going forward.  
\subsubsection{Step 1.b:  Maximum allowable multi-frequency power variation at SG terminal} The sum of frequency components in the SG terminal power should be less than the lowest value of the curve shown in Fig. \ref{fig:4mac_isolated} for each SG (see \eqref{eq:SumPe}), which are $1.17$ \% and $1.02$ \% of G3 and G4 ratings, respectively.
\subsubsection{Step 2: Calculation of algebraic IFs} %Dynamic model of network is required to determine the ground truth IF which is a frequency dependent quantity. We propose the calculation of algebraic IF which is sufficient to determine $P_{dc}^{max(j)}$ from $P_e^{max(i)}$. Fig. \ref{fig:comp_IF_4mac} shows the accuracy of algebraic IF compared to the ground truth IF. 
Table \ref{tab:4mac_IF} shows a comparison of the maximum allowed AI DC load fluctuation ($P_{dc,actual}^{max(j)}$) obtained from the transfer functions between the shaft section stresses and the data center power using the DP model with that obtained using algebraic IF ($P_{dc}^{max(j)}$). We see that the algebraic IFs provide a conservative value of $P_{dc}^{max(j)}$ for bus $9$ and an over-conservative value for bus $7$ that can be attributed to the placement of the DC near two GFL IBRs, which are considered as PV buses instead of slack buses during load flow in Steps $2.3$ and $2.4$. %One interesting observation is that a longer electrical distance from the SG may not necessarily imply a larger allowable AI DC load fluctuation. In this case, that trend is correctly captured in algebraic IFs.

\subsubsection{Step 3: Determine optimal allowable fluctuations in presence of multiple AI DC loads} 
For this case study, we considered that AI DC is located only at bus $9$ after screening out bus $7$ due to relatively low $P_{dc}^{max(j)}$. For bus $9$, we have $P_{dc}^{max(j)} = \min \limits_i {\frac{P_e^{max(i)}}{IF_{ij}}}$ = $32.39$ MW. 

\subsubsection{Time-domain validation using framework in section~\ref{sec:Framework}} Figure \ref{fig:4mac_time_domain} (a) shows the AI DC load power in bus $9$. The result of FFT on a $10$ s window (see, \textit{Remark} \ref{rem:3}) of $P_{dc}$ in Fig.~\ref{fig:4mac_time_domain} (b) shows that there is a $20$ Hz component (deliberately chosen to coincide with a torsional mode frequency in Fig.~\ref{fig:4mac_isolated} to simulate worst case) with $32.39$ MW amplitude in the subsynchronous range, which is equal to $P_{dc}^{max}$. \textit{Therefore, the load fluctuation should be safe based on our proposed approach.}

It can be seen from Fig.~\ref{fig:4mac_time_domain} (c) that the resulting p.u. stress (cyan) in the G4 generator-exciter shaft section is below the maximum allowable value (orange) obtained from Fig.~\ref{fig:Goodman} and taken as base. Note that this is the worst case stress amplitude among the shaft sections in G4. The frequency deviation (cyan) of G4 is also below the maximum allowed (orange) value (Fig.~\ref{fig:4mac_time_domain} (d)). \textit{This is in line with our expectation from the proposed approach.}

\begin{figure}
	%\vspace{-5pt
	\centering
	\includegraphics[width= 0.4\textwidth]{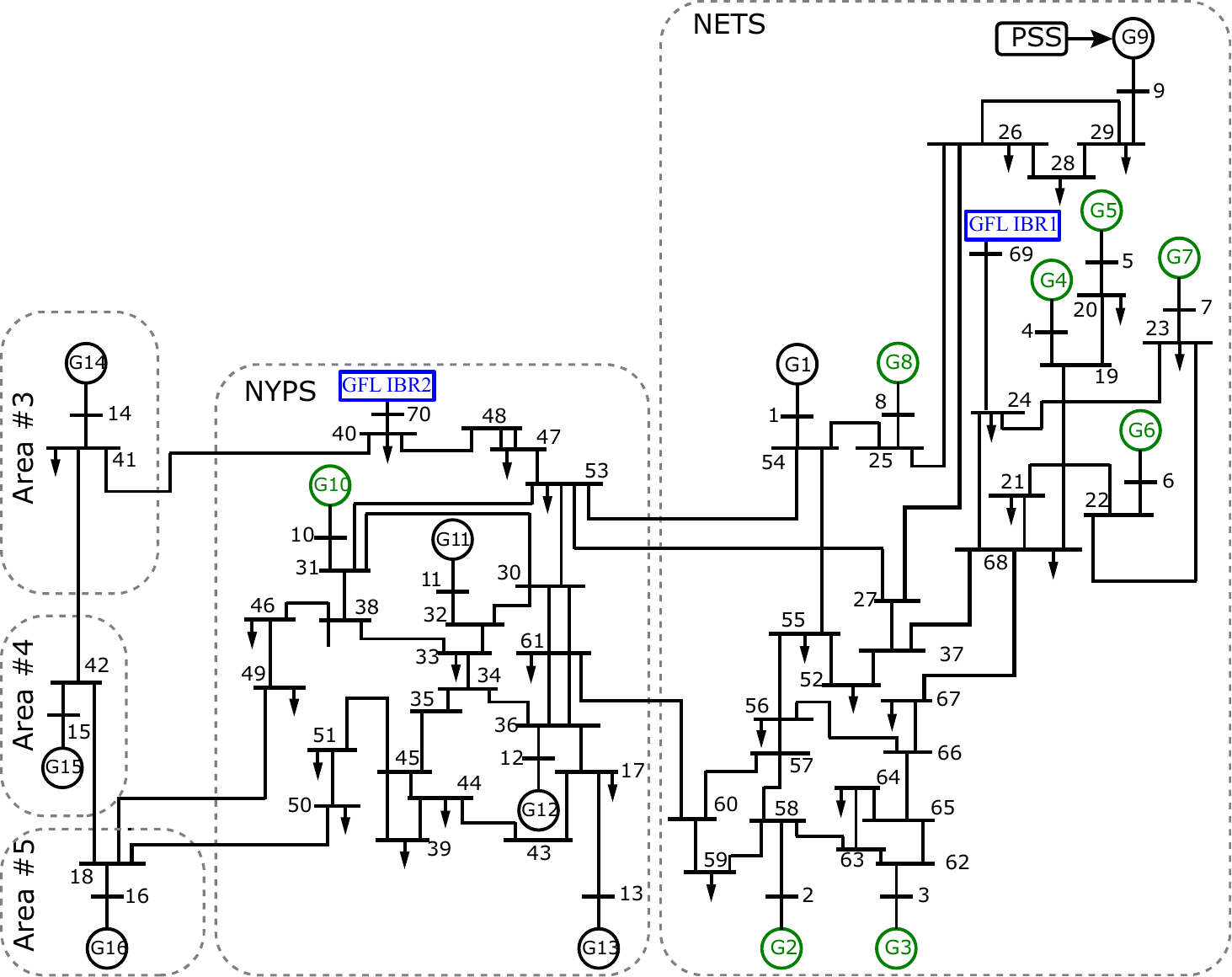}
	 \vspace{-5pt}
	\caption{\small{Modified IEEE-68 bus system \cite{Ameli-25-TPWRS}.}}
	\label{fig:IEEE-68 bus system}
	\vspace{-10pt}
\end{figure}

% \begin{figure}
% 	%\vspace{-10pt}
% 	\centering
% 	\includegraphics[trim = {1.8cm 7.5cm 2cm 8cm}, clip,width= 0.4\textwidth]{Figures/IF_matrix_4mac.pdf}
% 	 \vspace{-10pt}
% 	\caption{Step 2 for IEEE 4-machine system.}
% 	\label{fig:dp emt validation}
% 	\vspace{-7.5pt}
% \end{figure}

\vspace{-10pt}
\subsection{Application on the Modified IEEE-68 bus System}\label{sec:68bus}
Figure \ref{fig:IEEE-68 bus system} shows the modified IEEE-68 bus system where two GFL IBRs are connected to buses $24$ and $40$. Generators G2-G8 and G10 (indexed as $i = 1, \cdots, 8$) are modeled with multi-mass turbines. Generator G1 is a hydro unit and the local mode of G9 is stabilized through a PSS, which we did not wish to redesign in presence of the multi-mass model. 
% Therefore, G1 and G9 are modeled with single-mass rotors. 
All the remaining generators have aggregated models or representing areas. Hence, they are not modeled considering multi-mass turbine. Multi-mass turbine parameters for (G3, G4, and G6) and (G2, G5, G7, G8, and G10) are taken from \cite{IEEESSRBenchmark1977} and \cite{kundur1994power}, respectively.

\begin{figure}[!t]
	\vspace{-5pt}
	\centering
	\includegraphics[trim = {1.4cm 7.5cm 2cm 8.2cm}, clip,width= 0.45\textwidth]{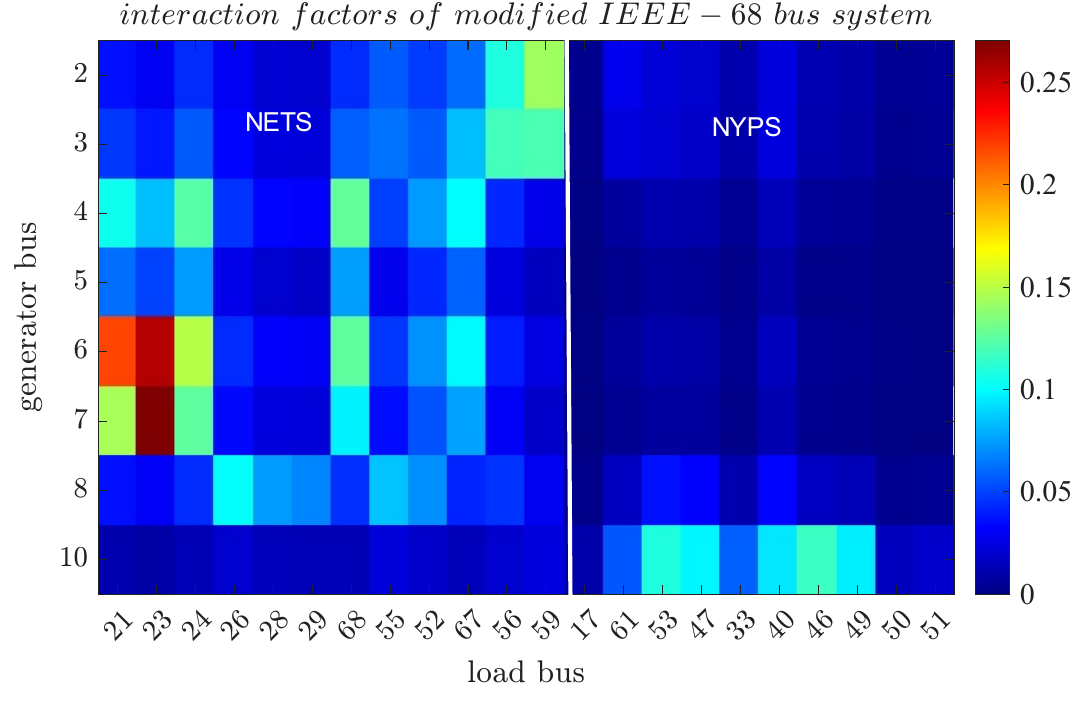}
	 \vspace{-15pt}
	\caption{\small{Algebraic IFs of modified IEEE-68 bus system.}}
	\label{fig:16mac_heatmap}
	\vspace{-17pt}
\end{figure}

After calculating the maximum allowable multi-frequency power variations at eight SG terminals, the algebraic IFs are evaluated. Figure \ref{fig:16mac_heatmap} shows the algebraic IFs between all possible AI DCs buses and SGs equipped with multi-mass turbines in the form of a heatmap that is self explanatory. Low IFs are observed between load buses in NYPS and generator buses in NETS, which is intuitive. \textit{This highlights that one may consider only areas of reasonable size for this analysis and not the whole system;} therefore, we consider only the buses in NETS as candidates, which are ranked based on p.u. $P_{dc}^{max(j)}$ in Table \ref{tab:16mac_optimization}. The values of $P_{dc}^{max(j)}$ obtained from algebraic IFs are conservative and their ratios with $P_{dc, actual}^{max(j)}$ range between $0.75$ and $1$.
%Figures \ref{fig:IF_bode_21} and \ref{fig:IF_bode_56} demonstrates that algebraic IF matches well with ground truth IF on average. 
% Figure \ref{fig:16mac_barplot} shows the ratio of the maximum allowable power variations in buses in NETS using our approach and the actual values obtained from the DP model. The ratios indicate that the algebraic IFs are reasonably conservative.  
\begin{table}[!t]
\caption{\small{Ranking AI DC buses based on $P_{dc}^{max(j)}$ in MW}}
\label{tab:16mac_optimization}
\centering
\begin{tabular}{|c|c|c|c|c|c|c|}
\hline
\textbf{Bus} & 56 & 68 & 55 & 67 & 24 & 29 \\
\hline
\textbf{$P_{dc}^{max(j)}$} & 84 & 82 & 81 & 80 & 74 & 71 \\
\hline
\textbf{Bus} & 59 & 28 & 21 & 23 & 52 & 26 \\
\hline
\textbf{$P_{dc}^{max(j)}$} & 59 & 52 & 51 & 43 & 40 & 35 \\
\hline
\end{tabular}
\vspace{-7.5pt}
\end{table}

\begin{table}[!t]
\caption{\small{Solution of LP \eqref{eq:LP_1} in MW}}
\label{tab:16mac_optimization_12bus}
\centering
\begin{tabular}{|c|c|c|c|c|c|c|}
\hline
\textbf{Bus} & 56 & 68 & 55 & 67 & 24 & 29 \\
\hline
\textbf{$P_{dc}^{(j)}$} & 13 & 13 & 13 & 13 & 12 & 11 \\
\hline
\textbf{Bus} & 59 & 28 & 21 & 23 & 52 & 26 \\
\hline
\textbf{$P_{dc}^{(j)}$} & 12 & 8 & 8 & 7 & 6 & 6 \\
\hline
\end{tabular}
\vspace{-5pt}
\end{table}

\begin{figure}[!t]
    \centering

    % First row
    \begin{subfigure}{0.24\textwidth}
        \centering
        \includegraphics[trim={0.8cm 7.5cm 1cm 8.2cm}, clip, width=\textwidth]{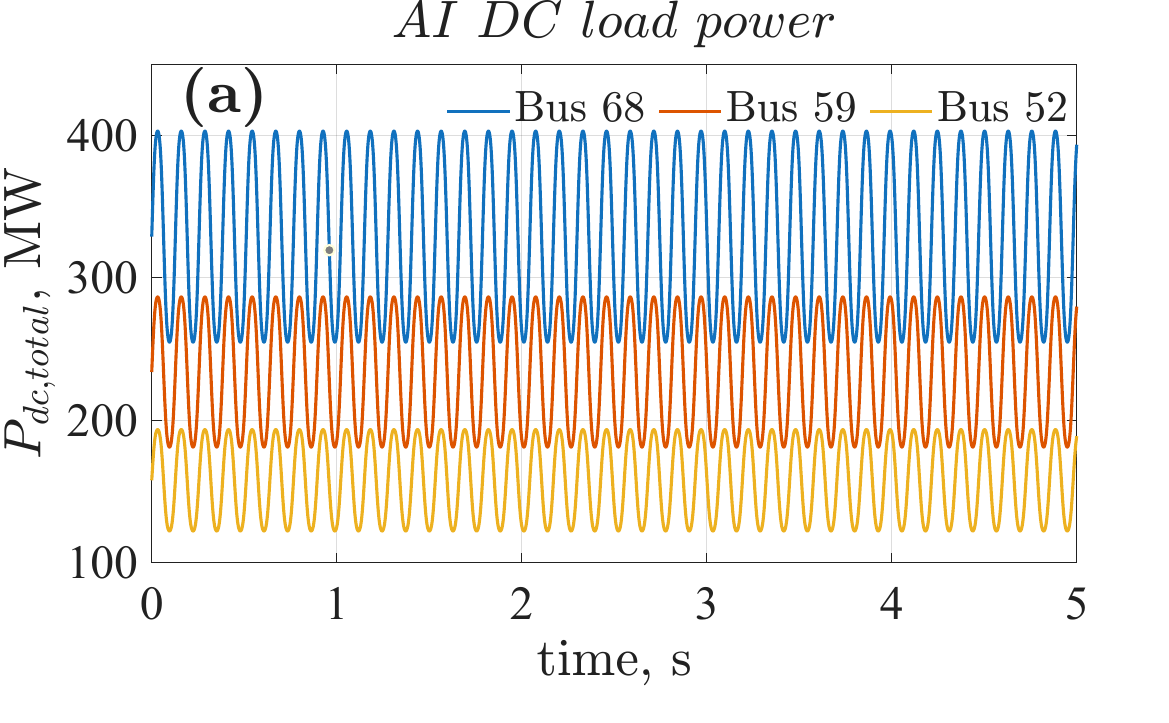}
        \vspace{-19pt}
        % \caption{}
        \label{fig:sub1}
    \end{subfigure}
    \hfill
    \begin{subfigure}{0.24\textwidth}
        \centering
        \includegraphics[trim={0.8cm 7.5cm 1cm 7cm}, clip, width=\textwidth]{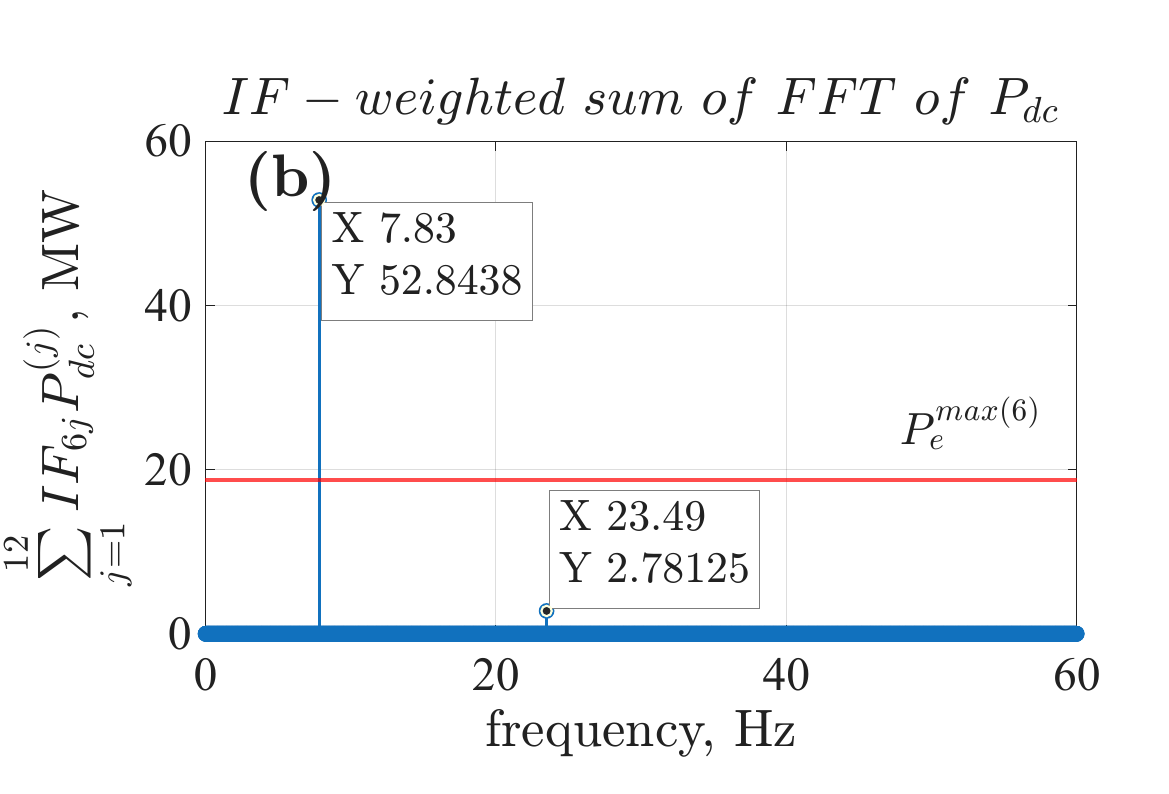}
        \vspace{-19pt}
        % \caption{}
        \label{fig:sub2}
    \end{subfigure}

    % Second row
    \begin{subfigure}{0.24\textwidth}
        \centering
        \includegraphics[trim={0.8cm 7.5cm 1cm 7.5cm}, clip, width=\textwidth]{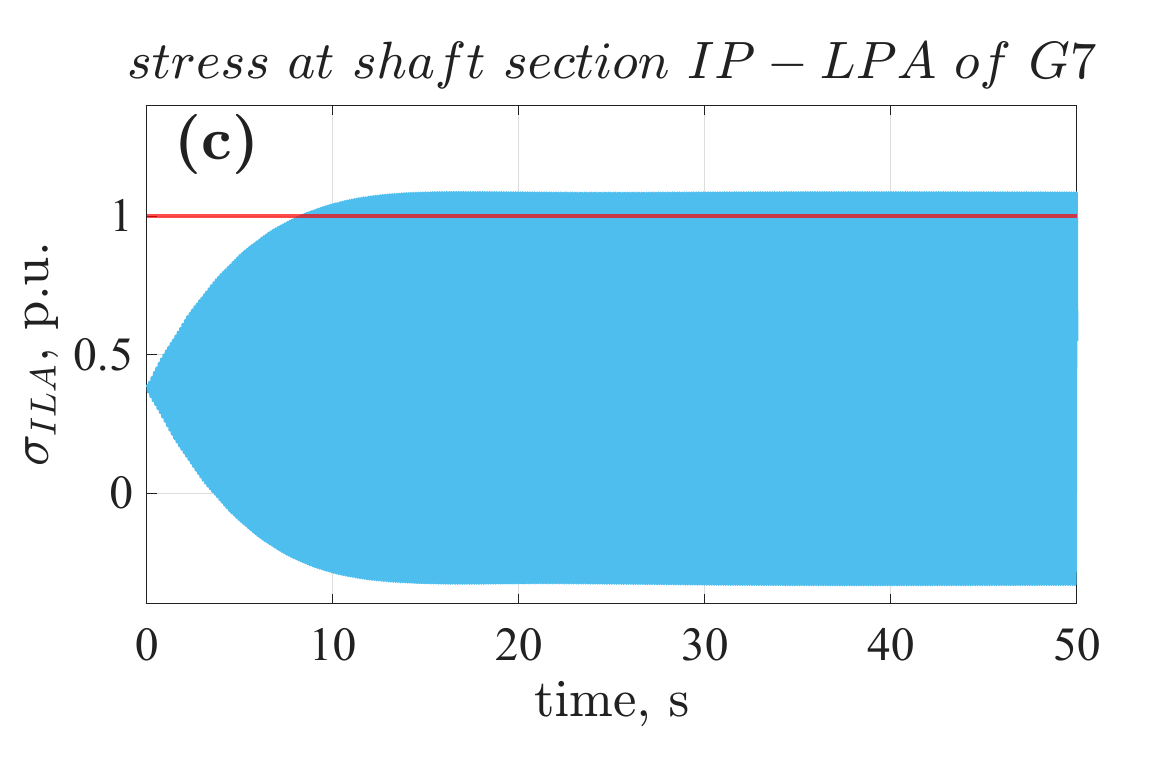}
        \vspace{-20pt}
        % \caption{}
        \label{fig:sub3}
    \end{subfigure}
    \hfill
    \begin{subfigure}{0.24\textwidth}
        \centering
        \includegraphics[trim={0.8cm 7.5cm 1cm 7.5cm}, clip, width=\textwidth]{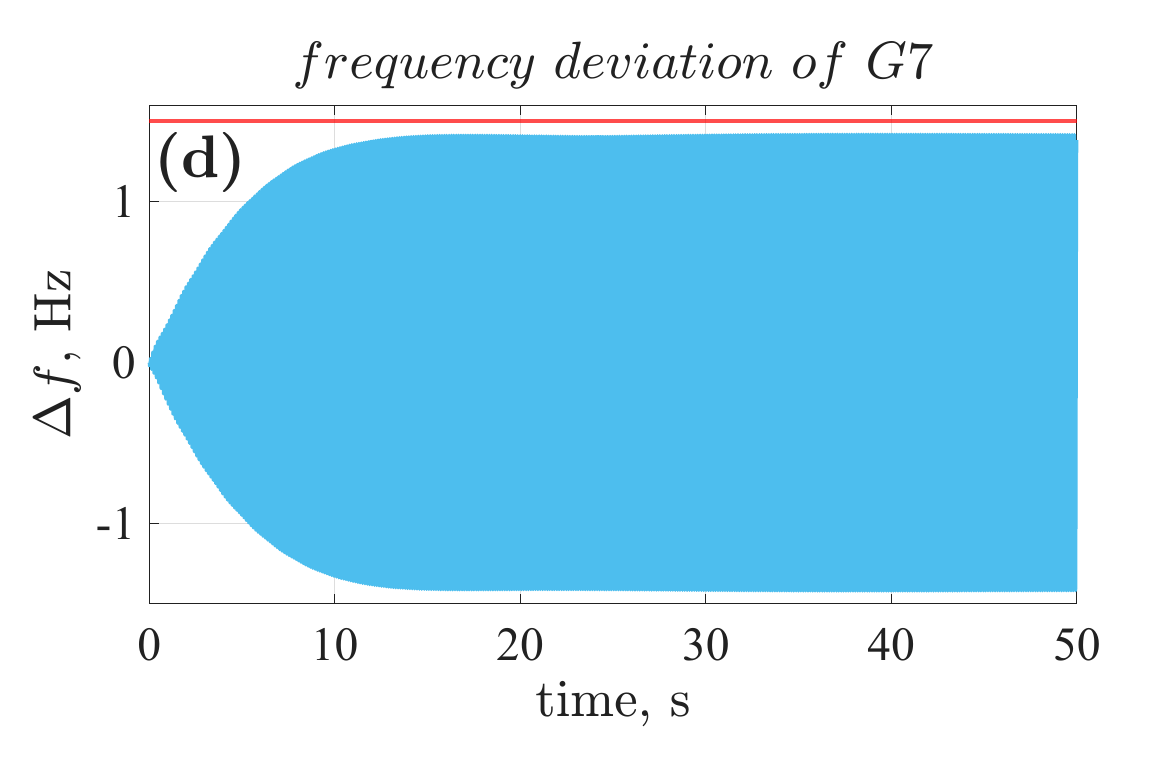}
        \vspace{-20pt}
        % \caption{}
        \label{fig:sub4}
    \end{subfigure}

    \caption{\small{AI DC load variation equaling $P_{dc}^{max(j)}$ with twelve NETS candidate buses: (a) AI DC load variation in three representative buses; (b) $\sum_{j=1}^{12} IF_{6j}P_{dc}^{(j)}$ overlapped with $P_{e}^{max(6)}$ ; (c) p.u. stress (cyan) in shaft section IP-LPA of G7 with maximum allowable value (orange) as base; and (d) frequency deviation (cyan) overlapped with maximum allowed (orange).}}
    \label{fig:16mac_stress_break}
    \vspace{-10pt}
\end{figure}

Suppose that we consider all of the twelve buses with AI DC loads in NETS (since the lowest value in Table \ref{tab:16mac_optimization} is $35$ MW, which is not very small) and we allow variations of $P_{dc}^{max(j)}$ on each of these buses. Figure~\ref{fig:16mac_stress_break}(a) shows the AI DC load powers in three representative buses among the twelve. Results from FFT on these variations reveal a $7.8$ Hz (torsional mode of G7 deliberately chosen for worst case) and a $23.4$ Hz component. The weighted sum $\sum_{j=1}^{12} IF_{6j}P_{dc}^{(j)}$ of each component corresponding to $i = 6$, i.e., G7 is shown in Fig.~\ref{fig:16mac_stress_break}(b). We observe that the aggregate of these two weighted sums is equal to $55.62$ MW, which is greater than the maximum allowed limit at the G7 terminal $P_e^{max(6)}$, which is equal to $18.76$ MW that in turn violates one of the first set of inequality constraints in \eqref{eq:LP_1}. This results in a stress in the IP-LPA shaft section of $G7$ that crosses and permanently stays above the allowed limit, as observed from Fig.~\ref{fig:16mac_stress_break}(c). Applying the \textit{Rainflow counting algorithm} \cite{ASTM_E1049_1985R2017} followed by \textit{Palmgren-Miner rule} \cite{Miner}, it can be determined that the shaft section will break. This example shows the necessity of solving the LP problem for multiple AI DC loads. 

\begin{figure}
	\vspace{-0pt}
	\centering
	\includegraphics[trim = {0.8cm 7cm 2cm 8.1cm}, clip,width= 0.45\textwidth]{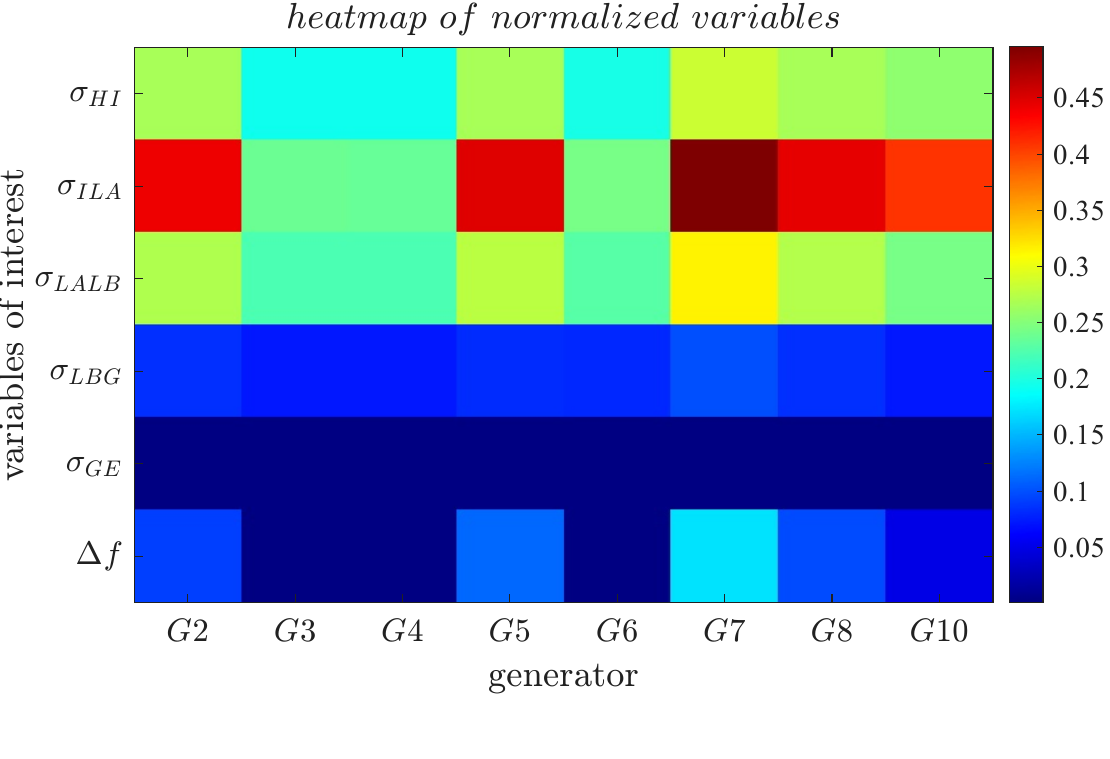}
	 \vspace{-17pt}
	\caption{\small{Fluctuations in AI DC loads respect limits obtained from LP: Heatmap of the resulting  transient peak stresses and deviations in frequency normalized with respect to corresponding limits in eight SGs in modified IEEE-68 bus system.}}
	\label{fig:16mac_heatmap_variables}
	\vspace{-15pt}
\end{figure}

The optimal values from the solution of the LP are shown in Table \ref{tab:16mac_optimization_12bus}. When these optimal variations are respected, the resulting \textit{transient peak} stresses in each shaft section of the eight SGs and \textit{transient peak} frequency deviations of those SGs remain within the safe limits as shown in Fig. \ref{fig:16mac_heatmap_variables}.

\vspace{-15pt}

% \vspace{-5pt}
\subsection{Application on Synthetic $2000$-bus Texas System}
\begin{figure}
	%\vspace{-10pt}
	\centering
	\includegraphics[trim = {1.1cm 5.6cm 2.4cm 6.2cm}, clip,width= 0.44\textwidth]{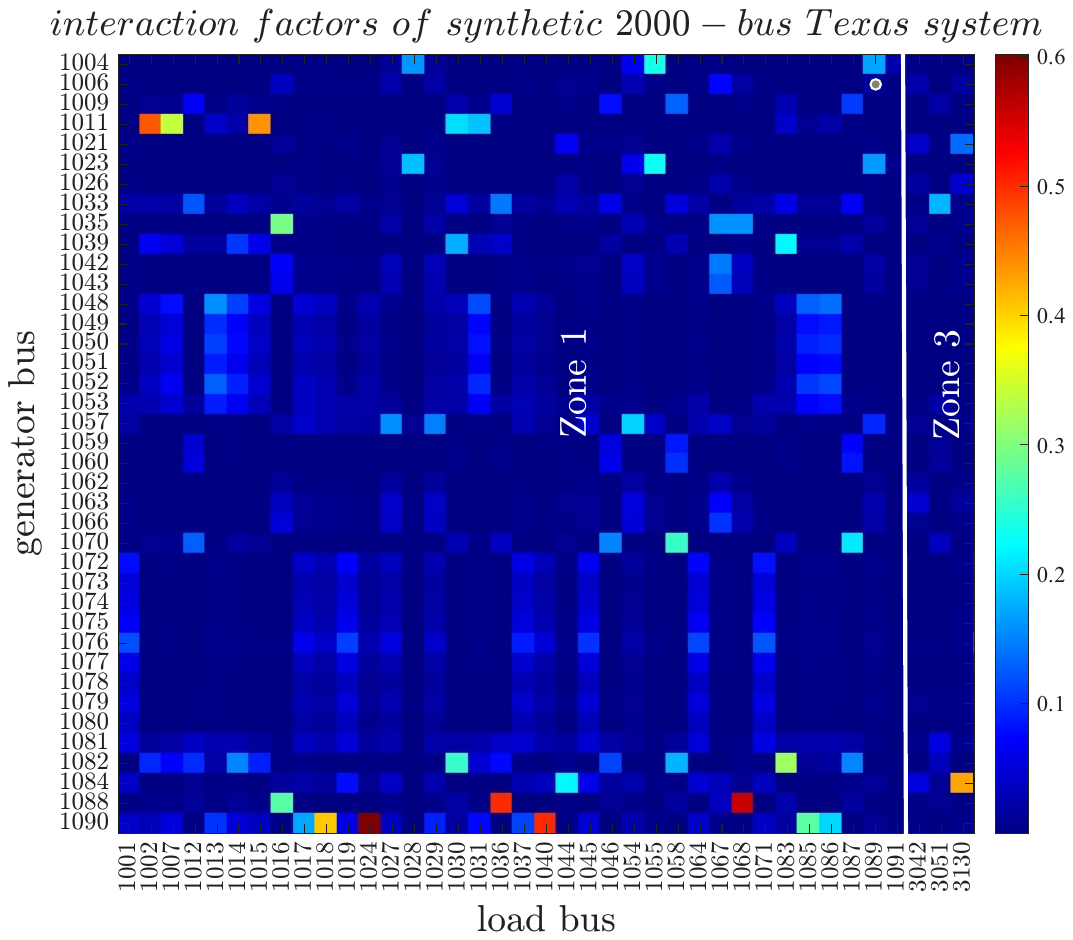}
	 \vspace{-10pt}
	\caption{\small{Algebraic IFs of 2000-bus Texas system.}}
	\label{fig:Texas_IF}
	\vspace{-7.5pt}
\end{figure}

\begin{figure}
	\vspace{-9pt}
	\centering
	\includegraphics[trim = {0.8cm 7.8cm 2cm 7.7cm}, clip,width= 0.44\textwidth]{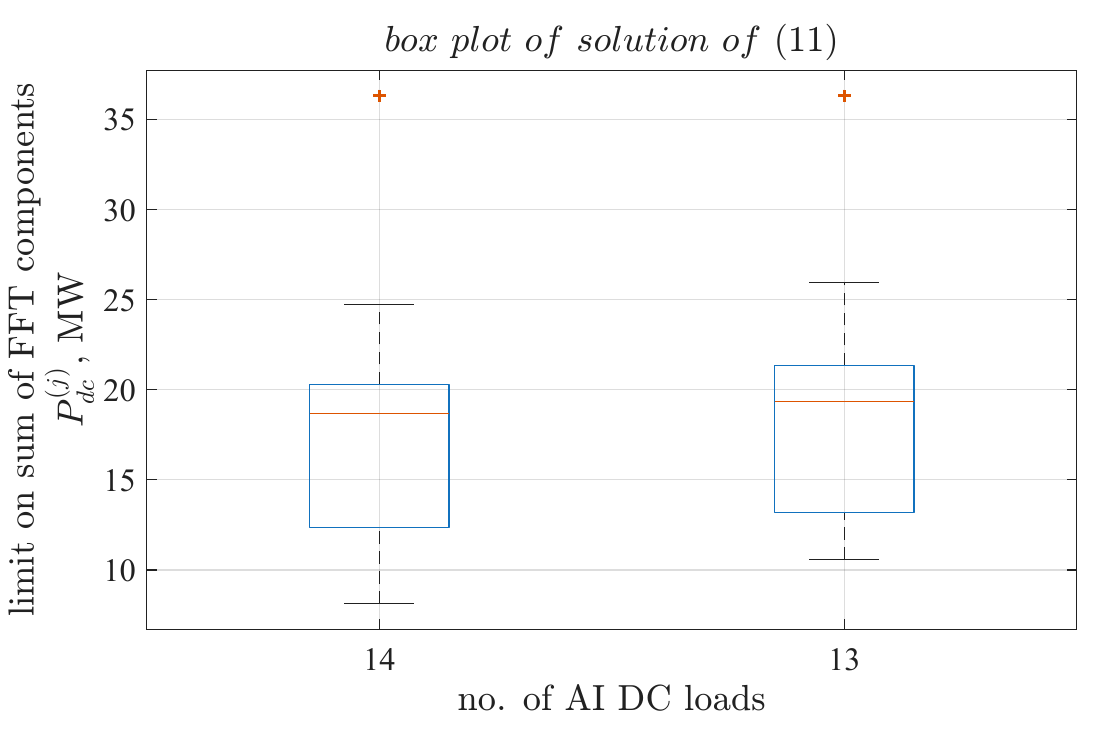}
	 \vspace{-10pt}
	\caption{\small{Box plots of LP solutions for 2000-bus Texas system.}}
	\label{fig:Texas_boxplot}
	\vspace{-19pt}
\end{figure}

To demonstrate the scalability of the proposed approach, the synthetic $2000$-bus Texas system with $485$ generators and $1123$ loads is considered. We solve the LP focusing on the $39$ generators in Zone $1$ assumed as thermal units ($P_e^{max(i)}$ assumed as $1\%$ of generator rating) and take all load buses in Zone $1$, as well as the adjacent load buses of the generator transformers as candidate buses. The heatmap of algebraic IFs are shown in Fig. \ref{fig:Texas_IF}, \textit{which demonstrates the applicability of our approach for a large-scale system.}  Through bus screening and ranking, we discarded all but $14$ of the $39$ buses where $P_{dc}^{max(j)}$ is less than $10$ MW (user-defined threshold).  LP optimization is run for the remaining $14$ load buses and it is found that for $1$ load bus $P_{dc}^{(j)}< 10~MW$ as shown in Fig. \ref{fig:Texas_boxplot}. Finally, we obtained $P_{dc}^{(j)}> 10~MW$ for the remaining $13$ load buses as shown in the boxplot of LP solutions.

\vspace{-5pt}
\section{Conclusions}\label{sec:Conclusions}
Our proposed framework assesses the impact of fluctuating AI DC loads on fatigue life of the shafts of thermal turbine-generators in the subsynchronous frequency range. The framework is based on the lumped multi-mass model of the mechanical side of the SGs and leverages the augmented modified Goodman diagram to determine the maximum allowable tensile stresses in shafts to avoid yielding or loss of fatigue life. We build on this framework to propose a scalable and easily accessible approach that helps to determine reasonably conservative optimal variations in AI DC loads in the frequency domain without impacting turbine fatigue life. The optimum limit is on the sum of the amplitudes of the subsynchronous frequency components of each AI DC load. An FFT can be performed on the AI DC power consumption to capture the frequency components and ensure compliance. Our analysis also reveals that an appropriate study area is sufficient for determining such limits without considering the whole system.      

\vspace{-8pt}
\bibliographystyle{IEEEtran}
\bibliography{Mybib}

\end{document}